%% file: ITSC2021.tex
\documentclass[letterpaper, 10 pt, conference]{ieeeconf}  

\usepackage{cite}
\usepackage{amsmath,amssymb,amsfonts}
\usepackage{algorithmic}
\usepackage{graphicx}
\usepackage{subcaption}
\usepackage{textcomp}
\usepackage{xcolor}
\usepackage{epstopdf}
\usepackage{tikz}
\usepackage{pgfplots}

\usetikzlibrary{patterns}

\newlength\figureheight
\newlength\figurewidth

\graphicspath{ {./images/} }

\DeclareMathOperator*{\argmin}{arg\,min}
\newcommand{\norm}[1]{\left\lVert#1\right\rVert}

\IEEEoverridecommandlockouts                              

\title{\LARGE \bf
	Safe Imitation Learning on Real-Life Highway Data for Human-like Autonomous Driving
}

\author{Flavia Sofia Acerbo{\textsuperscript {\tiny 1}}, Mohsen Alirezaei{\textsuperscript {\tiny 2}}, Herman Van der Auweraer{\textsuperscript {\tiny 1}}, Tong Duy Son{\textsuperscript {\tiny 1}} %
	\thanks{{\textsuperscript {\tiny 1}} Siemens Digital Industries Software, 3001 Leuven, Belgium
		{\tt\small flavia.acerbo@siemens.com}}%
	\thanks{{\textsuperscript {\tiny 2}} Siemens Industry Software and Services B.V., 5708 JZ Helmond, The Netherlands
		{\tt\small mohsen.alirezaei@siemens.com}}%
}

\begin{document}

\maketitle
\thispagestyle{empty}
\pagestyle{empty}

\begin{abstract}
This paper presents a safe imitation learning approach for autonomous vehicle driving, with attention on real-life human driving data and experimental validation. In order to increase occupant's acceptance and gain drivers’ trust, the autonomous driving function needs to provide a both safe and comfortable behavior such as risk-free and naturalistic driving. Our goal is to obtain such behavior via imitation learning of a planning policy from human driving data. In particular, we propose to incorporate barrier functions and smooth spline-based motion parametrization in the training loss function. The advantage is twofold: improving safety of the learning algorithm, while reducing the amount of needed training data. Moreover, the behavior is learned from highway driving data, which is collected consistently by a human driver and then processed towards a specific driving scenario. For development validation, a digital twin of the real test vehicle, sensors, and traffic scenarios are reconstructed toward high-fidelity and physics-based modeling technologies. These models are imported to simulation tools and co-simulated with the proposed algorithm for validation and further testing. Finally, we present experimental results and analyses, and compare with the conventional imitation learning technique (behavioral cloning) to justify the proposed development.
\end{abstract}

\section{Introduction}\label{intro}
Recently, {comfort} appears as a crucial challenge for automotive manufacturers and suppliers in Advanced Driver Assistance Systems (ADAS) development. The main objective is the design of ADAS functions that gain user acceptance and improve driving experience, while still providing an appropriate trade-off with safety requirements. The challenge includes not only algorithm developments but also finding appropriate testing and validation technologies. 

Since (good) human drivers are capable of operating vehicles in a comfortable manner, a way to improve autonomous driving control systems would be to recreate {human-like driving policies} in them \cite{inthepassengerseat}.
In this paper, we attempt to collect human driving data and mimic it with {imitation learning}. Although recent studies showed the potential of end-to-end learning \cite{nvidia, VISTA, Xiao_2020}, mapping images to controls, it is also known to have many disadvantages, such as sample inefficiency \cite{mobileyvse2e} and the lack of safety guarantees and stability properties, being prone to catastrophic errors even by low disturbances \cite{wu2021adversarial}. On the other hand, mid-to-mid approaches, combining learning and model-based components can provide interpretability of the framework and the possibilities to formally consider safety aspects \cite{chauffeurnet, pulver2021pilot, chen2019deep}.

In the proposed algorithm, we consider two main properties. First, safety should always be prioritized over comfort. Hence, constraints to avoid hazards on the passengers are explicitly treated and formally guaranteed. Secondly, sample efficiency is preferable. In the view of a novel design process of ADAS and Autonomous Vehicles (AV), involving continuous data collection, learning and updates throughout the entire life cycle of the product, we recognize the need of an agile learning, easy to integrate in an infinite development loop. Accordingly, it is  preferable that the designed learning modules are trained with small amounts of data and in a short amount of time. 

Recently, we have proposed \cite{ACC2020} a novel methodology for imitation learning combined with model-based information for training a neural network to output planned trajectories mimicking demonstrated ones. The methodology proposed to employ smooth spline-based motion planning in a loss-constrained neural network, using the convex hull property of B-splines. The approach has shown potential in increasing safety and sample efficiency of imitation learning algorithms (e.g. DAgger \cite{DAgger}). In this paper, the proposed safe imitation learning is validated on real-life data to learn a highway driving planning policy from humans.

Hence, the designed ADAS policies need to be properly tested to compare the learned behavior to the human collected one and validate their safety. Previous works, including ours, validated their algorithms in the same environment where the data was collected, e.g. generating simulated expert data from an optimal planner \cite{ACC2020, pulver2021pilot, chen2019deep}. However, in this case, human driving data is collected in the real world but, for safety and economical reasons, testing should be done in simulation. This induces the domain shift problem \cite{kelchtermans2018doshico}, i.e. the challenge of transferring a policy learned into the real domain to a simulated one. Indeed, following the learned trajectories in a simulated setting can produce a completely different distribution of feedback states, from vehicle and environment, and impair the performance evaluation of the algorithms. In this paper, we also demonstrate a testing approach to mitigate this challenge. It consists in the creation of a digital twin i.e., an accurate reconstruction of the real-life framework, made of all its components: vehicle dynamics, traffic scenario and perception.

The paper is organized as follows: Section \ref{background} provides background on imitation learning algorithms and on B-splines. Section \ref{il} formulates the safe imitation learning methodology. The description of the experimental setting and the validation results are provided in Section \ref{experiment}.

\section{Background}\label{background}
\subsection{Imitation Learning}
Imitation learning aims at learning how to mimic a certain behavior, called expert policy, defined as \(a^*_{t}=\pi^{*}(o_{t}, \theta^*)\), i.e. a mapping between observations and actions. A dataset of expert demonstrations is available \(D^{*}=\{(o^*_{1},a^*_{1}),...,(o^*_{N},a^*_{N})\}\), defining the set of visited states as \(S^{*}\). Applying supervised learning on \(D^{*}\), a policy \(\hat{\pi}\) is learned such that \[ \hat{\pi}=\argmin_{\pi\in \mathcal{C}}{\mathcal{L}}(\pi(\theta),\pi^{*}(\theta^*)), \] where $\mathcal{L}$ is usually represented by the Mean Squared Error between actions generated by the two policies and $\hat{\pi}$ is chosen to belong inside a class of functions $\mathcal{C}$, and with $\theta$ parameters. This is also known as {behavioral cloning} and the class $C$ is usually represented by neural networks. The learning is not always accurate, and \(\hat{\pi}\) will present imperfections that will lead the system, when in a closed-loop setting, to reach states not included in \(S^{*}\), for which the policy behavior becomes unpredictable. This is usually called {compounding of errors} problem and it is due to the fact that the statistical i.i.d. assumption between training and testing data is not valid for sequential predictions. Common solutions to mitigate the unsafe consequences of this problem require data augmentation. This can be done online by the expert, as in the DAgger algorithm \cite{DAgger}. However, this is impossible with a human expert, as in our case. It has also been proposed an entirely offline learning algorithm enriched by synthesized perturbations in the training dataset \cite{chauffeurnet}, creating {artificial corrective trajectories} from artificially perturbed states. The effect of adding such demonstrations, enhanced by having a small dataset, could be contamination in the driving pattern that we are trying to learn, acting as misguiding outliers. Hence, in this paper, both solutions are excluded. Instead, the integration of constraints and domain-aware information inside the training loss function of the neural network is proposed, through barrier function constraints.

\subsection{B-splines parametrization}
Splines are piecewise polynomial function that can represent trajectories as smooth, continuous functions by only a limited number of variables. The points where the pieces meet are called knots, which are sorted in non-decreasing order and not necessarily distinct. A spline can be expressed as a linear combination of B-splines basis function $B_{i}(\tau)$ with spline coefficients $\boldsymbol{\alpha}_{i}$ as
\begin{equation}
s(\tau)=\sum_{i=1}^{n} \boldsymbol{\alpha}_{i} B_{i}(\tau).
\end{equation}
The number of coefficients $n$ depends on the basis function degree $d$ and the number of spline knots $m$, that is, $n = m - d - 1$.
The coefficients \(\boldsymbol{\alpha}_{i}\) can be points in the \(x,y\) plane and are also called {control points}. 
A B-spline is always contained in the {convex hull} of its control points. 
Consequently, a constraint on the values of the \(\boldsymbol{\alpha}_{i}\) coefficients would imply an equivalent constraint on the amplitude of the spline function $s(\tau)$ for any $\tau$.  
Apart from imposing position constraints over the entire time horizon, we could also impose other kinematic limits such as velocity, acceleration and jerk constraints in the similar manner.

\section{Safe Imitation Learning}\label{il}
In our proposed safe imitation learning approach \cite{ACC2020}, the policy network makes use of {B-splines coefficients, i.e. points in 2D plane, to parametrize the output trajectory} and enhances the safety of the learned policy with domain-aware information, in the form of {barrier-function constraints added to the imitation loss}, keeping its trajectories contained in a safe convex hull.
Considering an imitation loss as \(\mathcal{L} = \norm{a_{t}-a_{t}^*}^2_2\), the constrained loss can be made by adding a barrier function, yielding: $\mathcal{L}_{safe} = \mathcal{L} + I(o_{t},a_{t},a_{t}^*)$.
The advantages of using B-splines lie mainly in the convex hull property. Secondarily, using coefficients as outputs allows a forced smooth trajectory, imposition of constraints at each time instant just by limiting the coefficient values and few output nodes. In this way, it is more scalable for longer horizons.

\begin{figure}
	\centering
	\includegraphics[width = 0.5\textwidth]{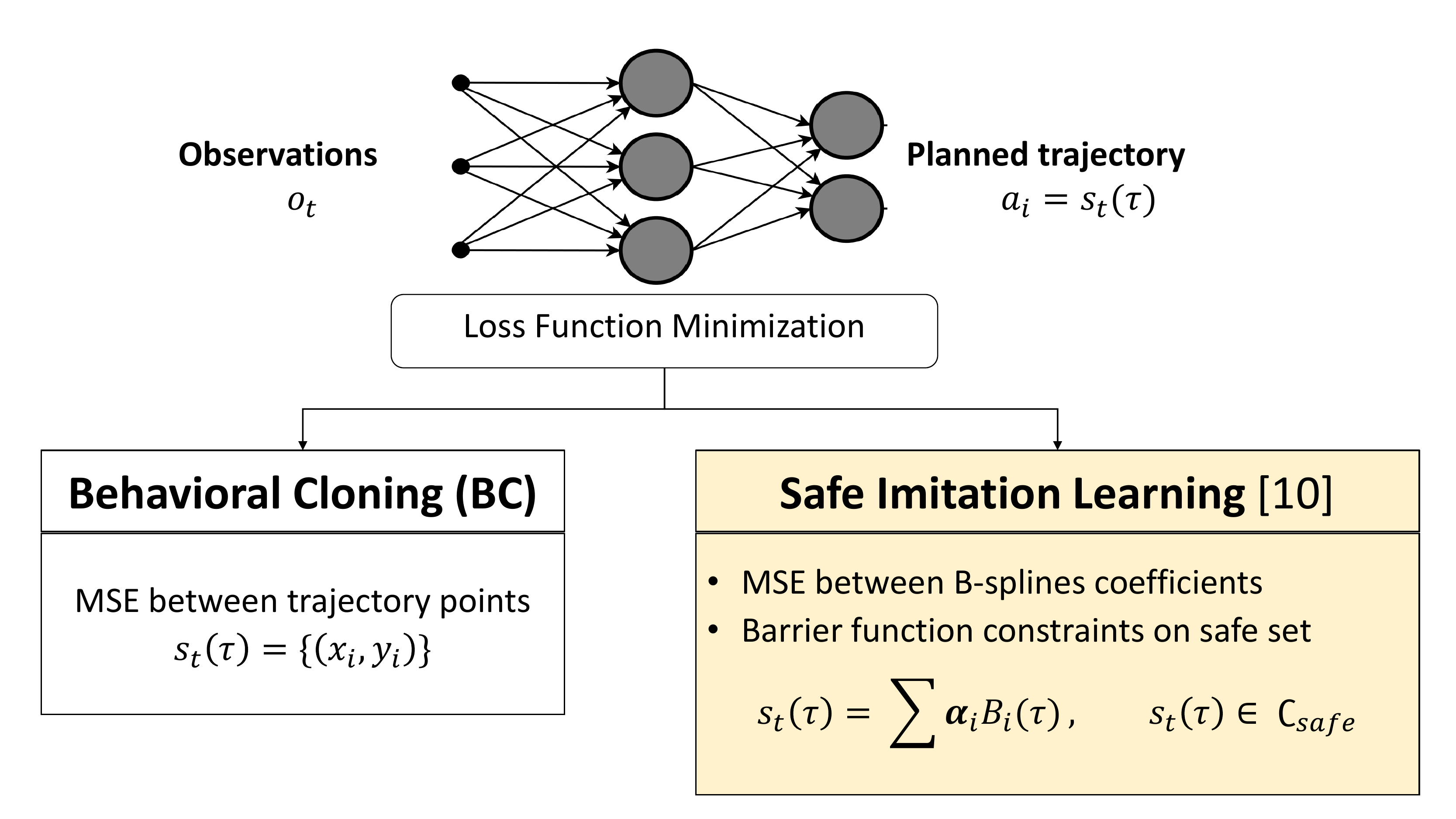}
	\caption{Safe Imitation Learning: the policy network outputs coefficients of B-splines (points in the 2D plane), and the loss function used during training is enhanced with barrier function constraints on the coefficient values, constraining the trajectory inside a safe convex hull. }
	\label{fig:SAFEIL}
\end{figure}

Eventually, our proposed learning method (as shown in Figure \ref{fig:SAFEIL}) can generate a driving imitation policy to avoid unsafe behaviors due to the compounding errors problem when tested in a closed-loop framework. The algorithms were trained and validated in simulators only.

\section{Experimental Validation on highway driving}\label{experiment}
The presented safe imitation learning algorithm is implemented and evaluated on human driving data in a highway scenario. Firstly, the data has been collected and processed, to be ready to use for imitation learning algorithms. Then, the planning and control system, including a policy network trained with safe imitation learning is implemented. Finally the designed system is validated with respect to its similarity with human driving data, and compared with a system trained with behavioral cloning, showing both ability in replicating the human driving pattern and in staying within the safety boundaries.
\subsection{Data Collection and Processing}\label{datacollproc}
\begin{figure}
	\centering
	\includegraphics[width = 0.48\textwidth]{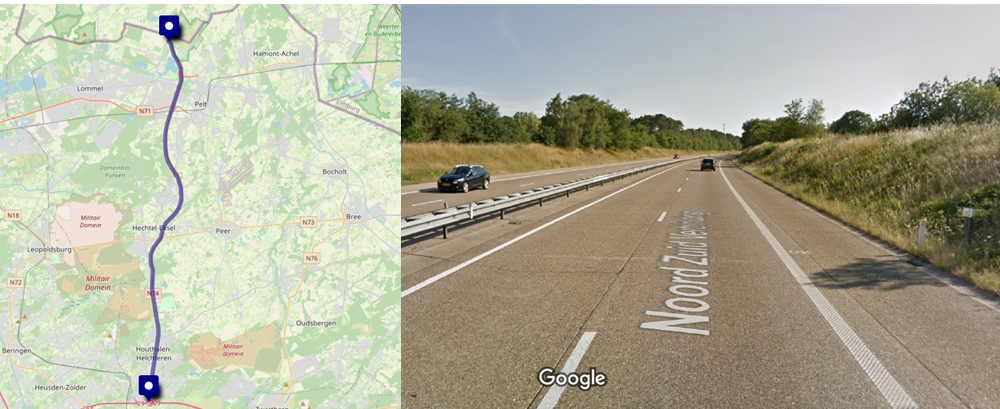}
	\caption{On the left, the road where the data was collected. The road is 29.2 kilometers long and it took around 30min to drive along it. On the right, a snapshot of how the road structure is like: highway-like, with low curvature and clear lane markers.}
	\label{fig:map}
\end{figure}

A Toyota Prius is equipped with different ADAS sensors and a user is asked to drive it naturally on \textit{N74/N715} roads (see Figure \ref{fig:map}).

\subsubsection{Vehicle, Sensors and Data Logging Architecture}\label{Vehicle Setup} 
\begin{figure*}
	\centering
	\begin{subfigure}[t]{0.3\textwidth}
		\centering
		\includegraphics[width=\textwidth]{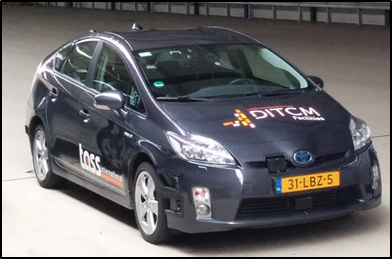}
		\caption{The vehicle}
		\label{fig:fig2a}
	\end{subfigure}
	\hfill
	\begin{subfigure}[t]{0.3\textwidth}
		\centering
		\includegraphics[width=\textwidth]{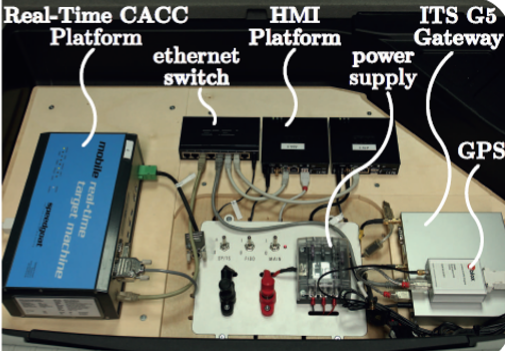}
		\caption{Close-up of the vehicle instrumentation
			in the trunk}
		\label{fig:fig2b}
	\end{subfigure}
	\hfill
	\begin{subfigure}[t]{0.3\textwidth}
		\centering
		\includegraphics[width=\textwidth]{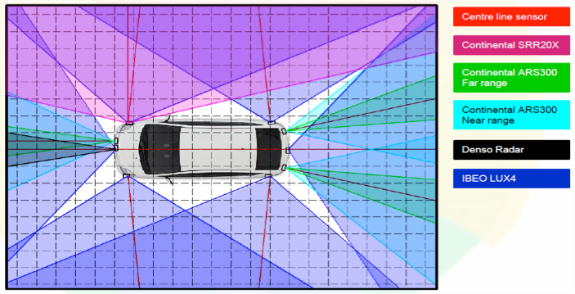}
		\caption{Sensors field of view}
		\label{fig:fig2c}
	\end{subfigure}
	\caption{Data Collection: vehicle, instrumentation and sensors}
	\label{fig:fig2}
\end{figure*}

To collect the data, a Toyota Prius (Figure \ref{fig:fig2}) has been used, retroﬁtted with the necessary equipment for data collection. Among others, it encompasses a Rapid Control Prototyping (RCP) system, ROS as a middleware for data logging, a modem for wireless communications (according to then ITS G5 standard), and a human-machine interface (HMI) computer platform to inform the driver via a dashboard screen. 
In particular, the Prius Carlab is equipped with a \textit{MobilEye} front camera, normal GPS, accurate positioning \textit{OxTS} system and 4 radars, of which two of them are fitted to the front and the other ones to the rear of the vehicle. Furthermore, the Carlab is equipped with an \textit{Ibeo} laser scanning system. The latter consists of 6 laser scanners, giving a 360 deg view around the vehicle with an angular resolution of 0.8 deg in the vertically, 0.125 deg horizontally, distance resolution of 4 cm, distance repetition accuracy of 10 cm, object classification (including velocity and direction information). Finally, it uses \textit{XSens} GPS and IMU to register ego vehicle position, motion and dynamics via the vehicle CAN.

\subsubsection{Data Processing}\label{dataproc}
\begin{figure}[b]
	\centering
	\resizebox{6cm}{!}{
	\input{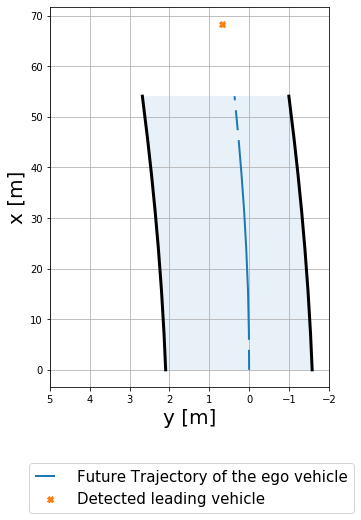}}
	\caption{Example of trajectory states and sensors information that can be extracted from the recorded human driving data.}
	\label{fig:fig1a}
\end{figure}

Firstly, a filtering is done by automatically removing all lane changes present in the data. Then, for each timestamp, the following entries are obtained:
\begin{itemize}
	\item X-Y coordinates: global coordinates in a reference frame with an origin that is zeroed every 5 minutes.
	\item Ego vehicle longitudinal speed.
	\item Coefficients of the fitting polynomials of the left and right lane boundaries, as in $y = c_0 + c_1x + c_2x^2 + c_3x^3$, where $y$ and $x$ are the lateral and longitudinal coordinates in the local vehicle reference frame.
	\item Leading vehicle coordinates: x-y position of the leading vehicle in the local vehicle reference frame, obtained from the radar.
\end{itemize}
An example of recorded trajectory extracted from these entries can be seen in Figure \ref{fig:fig1a}.

The task of autonomously driving on a defined lane on the highway involves two types of control: lateral control, which computes the steering wheel signal necessary to keep the vehicle on the lane, and longitudinal control, which computes the acceleration (throttle or brake) signal.
Highway longitudinal control is traditionally divided in {velocity control}, i.e. maintaining a desired speed without a leading vehicle in front, and {spacing control}, i.e. maintaining a desired distance from the leading vehicle. This driving control can be implicitly linked to human driving behavior style, although the desired speed and distance are not explicitly given by the driver. 
It is checked if these tasks are equally represented in the data: there is a clear prevalence of spacing control, i.e. most often, there is a leading vehicle that the driver has to keep the distance to. For these reasons, we only consider spacing control imitation learning and filter out the samples with no leading vehicle.

The leading vehicle speed can be computed by numerically differentiating two of its subsequent relative x-positions. However, we notice that many outliers are present, showing unfeasible leading vehicle speed values. By looking at the camera images recorded during those moments, these are happening in {cut-in/cut-out/merging} scenarios. Indeed, when a vehicle cuts in front of the ego vehicle, the radar will detect, in a single timestamp, an obstacle in front which is much closer than the previous one. This results in the erratic computation of an unfeasible deceleration of the leading vehicle. 

After filtering such scenarios, the dataset is divided in different \emph{car-following maneuvers}, in which the ego vehicle follows another one for the entire maneuver. Hence, the selected ADAS task can be seen as a combination of lane keeping and adaptive cruise control. From these maneuvers, experience tuples made of (current state, future trajectory) are created, where the future trajectory is obtained from the 20 samples ahead.

\subsection{Planning and Control System}
We propose to learn a planning policy, via our safe imitation learning approach, to perform the task of highway driving extracted in Section \ref{dataproc}. The policy takes the form of a shallow neural network, with manually chosen features. The policy, to plan a proper trajectory, shall take into account the direction of the road ahead and the detected distance and relative speed to the vehicle in front. Therefore, the inputs are chosen as: the coefficients of the second-order polynomial fitting the left and right lane lines, in the local reference frame of the ego vehicle, $\boldsymbol{c}_{l}$ and $\boldsymbol{c}_{r}$; the ego longitudinal vehicle speed, $v_x$; the leading vehicle speed, $v_{lead}$ and the distance from the leading vehicle, $d_{lead}$.

The network then outputs the planned trajectory $\boldsymbol{s}(t) = (x(t), y(t))$, for 20s ahead, which is then tracked by a low-level control. This is realized by a PID controller for the longitudinal dynamics and a Pure Pursuit for the lateral dynamics. The computed steering angle and acceleration signal, converted into throttle and brake controls, are given to the vehicle plant.
Specifically, a cubic spline ($d = 3$) with 8 knots is chosen (open uniform normalized knot vector $[0, 0 , 0, 0, 1, 1, 1, 1]$), resulting in $n = 4$ coefficients, each of them being a point in $\mathbb{R}^2$. Since the trajectory always starts from the origin of the current local reference frame, the coefficient multiplying the first B-spline is always null and can be neglected. Therefore, the total number of network outputs is 6, i.e. $\left(\alpha_{xi}, \alpha_{yi}\right)_{i=1,..,3}$, regardless of the length of the planned trajectory.
The barrier functions can be written as a linear combination of other functions $\sigma(z)$, in similar form to {softplus} functions: $\sigma(z) = ln(1 + e^z)$, limiting the values of the coefficients according to collision avoidance and lane boundaries constraints and, thanks to the convex hull property, the overall trajectory will be contained inside the safe area.

Here, the barrier function is selected as following: 
\begin{align}
& I(o_t,a_t)  \notag \\
& = K\sum_{i=1}^3{ \sigma\left(\alpha_{yi} - \left(c_{0l} + c_{1l}\alpha_{xi} + c_{2l}\alpha_{xi}^2 - \frac{T}{2}\right) \right)} \label{l1} \\
& + K\sum_{i=1}^3{ \sigma\left(-\alpha_{yi} + \left(c_{0r} + c_{1r}\alpha_{xi} + c_{2r}\alpha_{xi}^2 + \frac{T}{2}\right) \right)} \label{l2} \\
& + K\sum_{i=1}^3{ \sigma \left( \alpha_{xi} - d_{lead}(i) \right)},  \label{l3}
\end{align}
where $K = 1000$. For the lane boundaries constraints [(\ref{l1}), (\ref{l2})]: $T$ represents the track, i.e. the width, of the vehicle and $c_{il}$ and $c_{ir}$ are respectively the coefficients of the polynomials fitting the left and right lane. For the collision constraints (\ref{l3}): the leading vehicle is modeled with constant speed, such that $d_{lead}(i) = d_{lead}(t = 0) + v_{lead} t(i)$, where $t(i)$ is the time instant corresponding to the $i^{th}$ control point, computed from the knot vector.


\subsection{Closed-Loop Validation Methodology}
Also on-policy performance should be tested, i.e. letting the network act on the actual system, in closed-loop, and compare the resulting states with the ones recorded during human driving.

This is a challenge that can be addressed through the use of a {digital twin} \cite{Hartmann2020DigitalT}, i.e. a simulated replica of the scenario and platform where the human driving data is collected.


The digital twin that fits our use case is made of a virtual driving scenario, which will give the feedback inputs to the planner, and of a high-fidelity vehicle model, so that the provided control actions can generate realistic vehicle dynamics behaviors. This combination, one focused on the vehicle dynamics and one focused on the environment outside the vehicle, yields the basis for the virtual validation of the learned planner \cite{adasframeworktest}.

\subsubsection{Leave-One-Out and Scenario Creation in Simcenter Prescan\textsuperscript{\textcopyright}}
From the maneuvers extracted from the dataset, a sample maneuver can be extracted and excluded from the training data. For example, 50 samples of a 10 seconds car-following maneuver at 115km/h are left out. From the recorded dataset, at each timestamp, the position of the lane boundaries and the speed of the leading vehicle are known, as well as the initial states. This information is enough to easily recreate the scenario. To do this, Simcenter Prescan is employed: a traffic simulator which is also able to model ADAS sensors at different fidelity levels.  

\begin{figure}
	\centering
	\includegraphics[width = 0.48\textwidth]{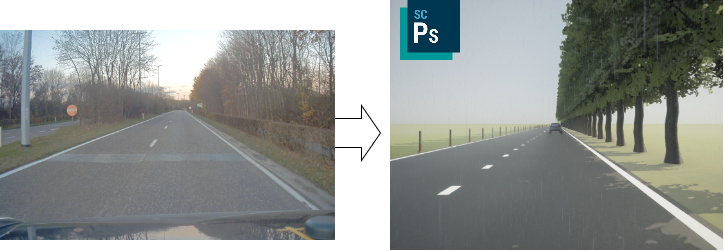}
	\caption{Recreation of the traffic scenario in Simcenter Prescan\textsuperscript{\textcopyright}}
	\label{fig:prescan}
\end{figure}

\subsubsection{Full Vehicle Modelling in Simcenter Amesim\textsuperscript{\textcopyright}}
A vehicle dynamics model for ADAS validation is not trivial to obtain: while Automotive OEMs have access to full or high-fidelity vehicle models, having such information is usually a significant challenge for software suppliers who need to develop and validate their ADAS solutions. In this project, a {test-driven full vehicle model identification} is proposed. The approach maximizes the use of vehicle driving data, while limiting as much as possible the need for bench tests. The identified vehicle characteristics are then easily plugged in a 15 degree-of-freedom model in Simcenter Amesim.

\subsection{Validation Results}
According to the presented strategy, the proposed planning and control system is validated. Two planning policies are trained: one using our safe imitation learning approach and one using standard behavioral cloning, as baseline. We evaluate the two resulting systems for: 
\begin{itemize}
	\item Human-like performance: we replicate a collected driving scenario in simulation and compare the closed-loop simulated states with human real-life ones.
	\item Safety performance: we define multiple artificial test scenarios, in simulation and with varying external parameters, and check the ability of the two controllers to respect safety constraints.
\end{itemize}

\begin{figure}
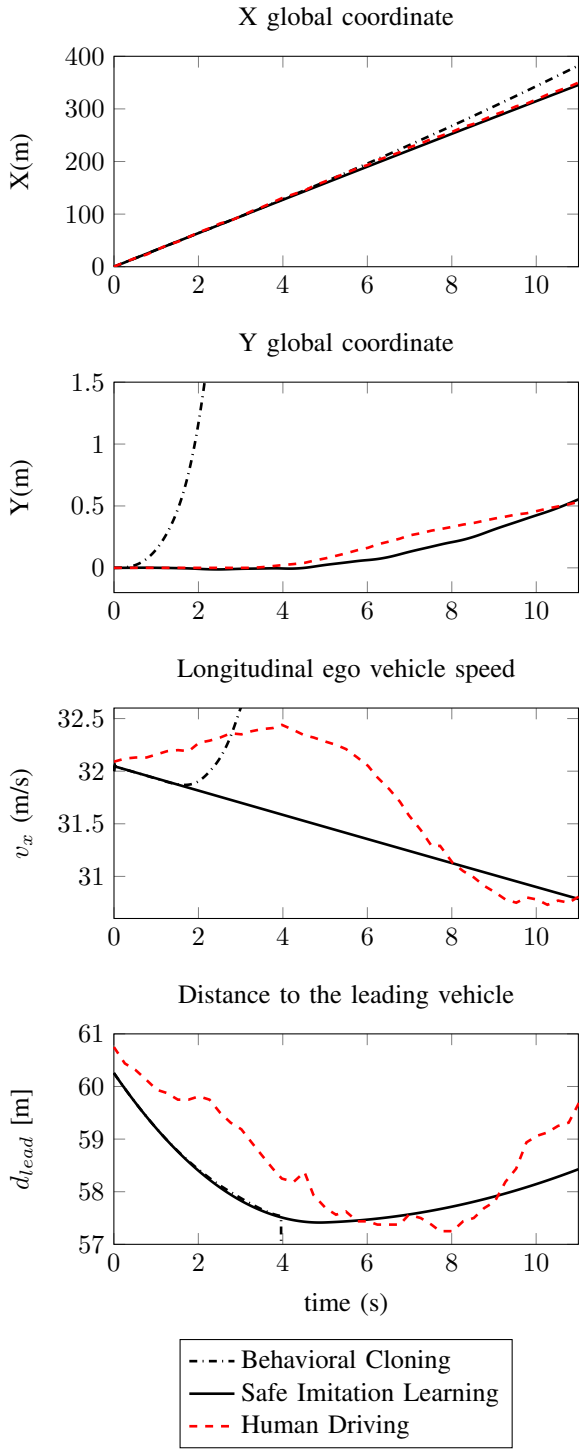

	\input{images/validate_xpos.tikz}
	\vskip .3cm	
	\input{images/validate_ypos.tex}
	\vskip .3cm	
	\input{images/validate_vx.tex}
	\vskip .3cm	
	\input{images/validate_dlead.tex}
	\caption{Human-like Perfomance Validation Results}
	\label{fig:validate}
\end{figure}

\paragraph{Human-like performance}
As shown in Figure \ref{fig:validate}, with behavioral cloning the compounding error problem causes the vehicle to drift off the lane after a few seconds. This makes it impossible to check the human-like performance of the policy. The descending step in the lower right diagram (\emph{Distance to the leading vehicle}) is due to sensors not detecting anymore the leading vehicle.
Conversely, by training the network with the safe imitation learning method, the vehicle can respect the lane boundaries also when deviating from the training features distribution.

From the figure, it is interesting to see how the planning network has an \emph{averaging effect} on the capture of the driving pattern: indeed, its main preferred behavior is replicated, but at the cost of losing the higher-frequency dynamics characterizing the collected human driving. This is expected, considering that the imitation loss considers a \emph{mean} of $L_2$ errors between different splines coefficients. 

\begin{figure}
	\input{images/safe}
	\caption{Safety Perfomance Validation Results: percentage of completed path before a safety constraint (lane boundaries or collision) is violated}
	\label{fig:safevalid}
\end{figure}
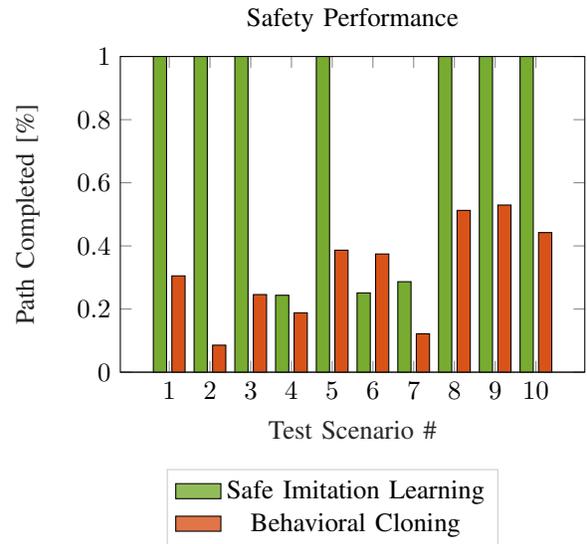

\paragraph{Safety performance}
Additionally, the two controllers are tested on other artificially created scenarios for car following. A curved track of 500m radius and 1km length is designed in Prescan, and a leading vehicle is added in front of the simulated ego vehicle, with a constant random speed between 25 and 32 m/s. Both algorithms are tested at the same time in closed-loop, and a flag is raised whenever a safety constraint is violated, i.e. when the vehicle exits its own lane or when it gets too close to the vehicle in front (time to collision $<= 1s$). Then, it is computed the percentage of completed path before the flag is raised, and the results can be seen in Figure \ref{fig:safevalid}. In general, safe imitation learning performs better, being able to complete paths, whereas behavioral cloning could not. In some scenarios (4-6-7), where safe imitation learning violated safety constraints, it can be seen that the leading vehicle was close to its possible minimum speed (25m/s), hence provided a smaller time-to-collision since the beginning of the path.

\section{Conclusions}
This paper presented the validation on real-life data of a safe imitation learning approach in the design of a human-like autonomous driving system for highway driving, during maneuvers concerning lane keeping and car following. For safety, sample efficiency and easy simulation-based testing, the system has been designed with a motion planning neural network policy followed by a model-based control. The planning policy has been learned to mimic the human driving patterns but, based on previous studies, has also been informed about safety constraints through barrier functions in the training loss, limiting the control points of a B-spline convex hull. The use of real-life data required a fine data processing, analytically discovering and removing the scenarios that acted as outliers in the learning process, e.g. cut-ins/cut-outs/merging. The system has been tested using a digital twin of a recorded real life scenario, where it was possible to compare the simulated vehicle states, controlled by our system, to the collected ones, controlled by the human driver. The results show that, in the tested scenario, the policy generates an averaged human-like driving behavior, presenting a similar pattern when considering slow-changing dynamics. Moreover, it has been shown that, without any possibility of online learning with interactive expert, our safe imitation learning is able to stay within safety limit, whereas a common behavioral cloning algorithm could not.

\section*{Acknowledgment}
This work has been conducted within the ID2CON (Integrated IDentification for CONtrol) and the H2020-EU FOCETA (Foundations for Continuous Engineering of Trustworthy Autonomy) projects. The authors would also like to thank Irfan M. Badshah, Ludovico Ruga and Lorenzo Lugo, from Siemens, for their supports in vehicle setting, data collection and full vehicle modelling.

\bibliographystyle{IEEEtran}
\bibliography{IEEEabrv,ITSC2021}
\end{document}

%% file: images/trajectory.tex
\begin{tikzpicture}

\definecolor{color0}{rgb}{0.12156862745098,0.466666666666667,0.705882352941177}
\definecolor{color1}{rgb}{1,0.498039215686275,0.0549019607843137}

\begin{axis}[
width=1\figurewidth,
height=3\figureheight,
legend cell align={center},
legend style={
  fill opacity=0.8,
  draw opacity=1,
  text opacity=1,
  at={(-0.1,-0.4)},
  anchor=south west,
  draw=white!80!black
},
tick align=outside,
tick pos=left,
x dir=reverse,
x grid style={white!69.0196078431373!black},
xlabel={y [m]},
xmajorgrids,
xmin=-2, xmax=3,
xtick style={color=black},
y grid style={white!69.0196078431373!black},
ylabel={x [m]},
ymajorgrids,
ymin=-3.41875, ymax=71.79375,
ytick style={color=black}
]
\path [fill=color0, fill opacity=0.05]
(axis cs:-1.58203125,0)
--(axis cs:2.09375,0)
--(axis cs:2.13513441253113,7.22469307666688)
--(axis cs:2.19857012801569,15.6590333402783)
--(axis cs:2.27742784770981,24.0623590886098)
--(axis cs:2.35711682791558,31.2730189982103)
--(axis cs:2.44764314018673,38.4921636306681)
--(axis cs:2.54844999302616,45.7039603426929)
--(axis cs:2.67827519209386,54.092823638719)
--(axis cs:-0.997506057906144,54.092823638719)
--(axis cs:-0.997506057906144,54.092823638719)
--(axis cs:-1.12733125697384,45.7039603426929)
--(axis cs:-1.22813810981327,38.4921636306681)
--(axis cs:-1.31866442208442,31.2730189982103)
--(axis cs:-1.39835340229019,24.0623590886098)
--(axis cs:-1.47721112198431,15.6590333402783)
--(axis cs:-1.54064683746887,7.22469307666688)
--(axis cs:-1.58203125,0)
--cycle;

\addplot [very thick, color0, dash pattern=on 10pt off 5pt on 100pt off 5pt]
table {%
0 0
0.00521638995451212 7.22469307666688
0.0274582111915151 15.6590333402783
0.0728469165887873 24.0623590886098
0.123843943092652 31.2730189982103
0.189337018990955 38.4921636306681
0.265185654873676 45.7039603426929
0.365734660257658 54.092823638719
};
\addlegendentry{Future Trajectory of the ego vehicle}
\addplot [very thick, black]
table {%
2.09375 0
2.13513441253113 7.22469307666688
2.19857012801569 15.6590333402783
2.27742784770981 24.0623590886098
2.35711682791558 31.2730189982103
2.44764314018673 38.4921636306681
2.54844999302616 45.7039603426929
2.67827519209386 54.092823638719
};
\addlegendentry{Lane Lines}
\addplot [very thick, black, forget plot]
table {%
-1.58203125 0
-1.54064683746887 7.22469307666688
-1.47721112198431 15.6590333402783
-1.39835340229019 24.0623590886098
-1.31866442208442 31.2730189982103
-1.22813810981327 38.4921636306681
-1.12733125697384 45.7039603426929
-0.997506057906144 54.092823638719
};
\addplot [thick, color1, mark=*, mark size=4, mark options={solid}, only marks]
table {%
0.6875 68.375
};
\addlegendentry{Detected leading vehicle}
\end{axis}

\end{tikzpicture}

%% file: images/validate_dlead.tex
\begin{tikzpicture}

\begin{axis}[%
width = 0.95\figurewidth,
height= \figureheight,
scale only axis,
xmin=0,
xmax=11,
xlabel={time (s)},
ymin=57,
ymax=61,
ylabel={$d_{lead}$ [m]},
title={Distance to the leading vehicle},
legend style={at={(0.5,-0.45)},anchor=north,legend cell align=left}
]
\addplot [color=black, dashdotted, line width=1.0pt]
  table[row sep=crcr]{%
0	60.2559940743791\\
0.01	60.2446184290356\\
0.02	60.2329004461424\\
0.03	60.2212231635804\\
0.04	60.2096211952586\\
0.05	60.1980195475708\\
0.06	60.1864476843456\\
0.07	60.1749071136516\\
0.08	60.1633876387338\\
0.09	60.1518937481938\\
0.1	60.1404257338134\\
0.11	60.1289834836001\\
0.12	60.1175681127107\\
0.13	60.1061796286784\\
0.14	60.0948178691096\\
0.15	60.0834828858311\\
0.16	60.0721745752041\\
0.17	60.0608928321167\\
0.18	60.0496376268743\\
0.19	60.038408983402\\
0.2	60.0272068487943\\
0.21	60.0160310790391\\
0.22	60.0048815095585\\
0.23	59.9937579763431\\
0.24	59.9826603258651\\
0.25	59.9715884143411\\
0.26	59.9605421041697\\
0.27	59.9495212556152\\
0.28	59.9385257052124\\
0.29	59.9275552125218\\
0.3	59.9166094842581\\
0.31	59.9056882511168\\
0.32	59.8947912699192\\
0.33	59.8839182969631\\
0.34	59.873069085703\\
0.35	59.8622434369372\\
0.36	59.8514411877654\\
0.37	59.8406622176836\\
0.38	59.8299064545095\\
0.39	59.8191738668248\\
0.4	59.8084644544554\\
0.41	59.7977782411679\\
0.42	59.7871152673402\\
0.43	59.7764755853752\\
0.44	59.7658592589997\\
0.45	59.755266360986\\
0.46	59.7446969698766\\
0.47	59.7341511678074\\
0.48	59.7236290377763\\
0.49	59.7131306614834\\
0.5	59.7026561196734\\
0.51	59.6922054925985\\
0.52	59.6817788603749\\
0.53	59.6713763030585\\
0.54	59.660997899264\\
0.55	59.6506437259784\\
0.56	59.6403138580655\\
0.57	59.6300083663008\\
0.58	59.6197273182649\\
0.59	59.6094707793407\\
0.6	59.5992388110327\\
0.61	59.5890314713913\\
0.62	59.5788488160539\\
0.63	59.5686908974131\\
0.64	59.5585577672343\\
0.65	59.5484494779878\\
0.66	59.538366080455\\
0.67	59.5283076232984\\
0.68	59.518274146767\\
0.69	59.5082656848227\\
0.7	59.4982822674242\\
0.71	59.4883239249028\\
0.72	59.4783906898888\\
0.73	59.4684825875095\\
0.74	59.4585996415011\\
0.75	59.4487418771314\\
0.76	59.4389093226411\\
0.77	59.4291020113357\\
0.78	59.419319977893\\
0.79	59.4095632561697\\
0.8	59.3998318805385\\
0.81	59.3901258834209\\
0.82	59.3804452955334\\
0.83	59.3707901492171\\
0.84	59.3611604776252\\
0.85	59.3515563204958\\
0.86	59.3419777273286\\
0.87	59.3324247457319\\
0.88	59.3228974153732\\
0.89	59.3133957720657\\
0.9	59.3039198478499\\
0.91	59.2944696729901\\
0.92	59.2850452801536\\
0.93	59.2756467021639\\
0.94	59.2662739727008\\
0.95	59.2569271291911\\
0.96	59.247606209458\\
0.97	59.2383112521939\\
0.98	59.2290422998194\\
0.99	59.2197993951715\\
1	59.2105825823041\\
1.01	59.2013919091332\\
1.02	59.1922274233954\\
1.03	59.1830891206686\\
1.04	59.1739770785348\\
1.05	59.1648913526541\\
1.06	59.1558319781003\\
1.07	59.146798991603\\
1.08	59.1377924285101\\
1.09	59.128812325002\\
1.1	59.1198587213312\\
1.11	59.1109316565752\\
1.12	59.102031165309\\
1.13	59.0931572774902\\
1.14	59.0843100158354\\
1.15	59.0754894017963\\
1.16	59.0666954693436\\
1.17	59.0579282910705\\
1.18	59.0491879896327\\
1.19	59.0404747241927\\
1.2	59.031788662805\\
1.21	59.0231299341974\\
1.22	59.0144985951981\\
1.23	59.0058946669269\\
1.24	58.9973181452514\\
1.25	58.9887689961936\\
1.26	58.9802472654455\\
1.27	58.9717529424627\\
1.28	58.9632860159881\\
1.29	58.9548464664819\\
1.3	58.9464342722698\\
1.31	58.938049423395\\
1.32	58.9296919173929\\
1.33	58.9213617629474\\
1.34	58.9130589876613\\
1.35	58.9047836277908\\
1.36	58.8965357156772\\
1.37	58.8883152684303\\
1.38	58.8801222774843\\
1.39	58.8719567217907\\
1.4	58.8638185836022\\
1.41	58.8557078414567\\
1.42	58.8476244762344\\
1.43	58.8395684817411\\
1.44	58.8315398504638\\
1.45	58.8235385673019\\
1.46	58.8155645655527\\
1.47	58.807617842006\\
1.48	58.79969837039\\
1.49	58.7918061136327\\
1.5	58.7839410320295\\
1.51	58.7761031195964\\
1.52	58.7682924354321\\
1.53	58.7605090733305\\
1.54	58.7527531481648\\
1.55	58.7450247953075\\
1.56	58.7373241354382\\
1.57	58.7296512471098\\
1.58	58.7220061678081\\
1.59	58.7143888893932\\
1.6	58.7067993847848\\
1.61	58.6992376321926\\
1.62	58.6917035924837\\
1.63	58.6841972243083\\
1.64	58.6767185082939\\
1.65	58.6692674163908\\
1.66	58.6618439229427\\
1.67	58.654448006708\\
1.68	58.6470796177995\\
1.69	58.6397387817864\\
1.7	58.6324255013356\\
1.71	58.6251397774337\\
1.72	58.6178816397879\\
1.73	58.610651165557\\
1.74	58.6034484072457\\
1.75	58.5962733974128\\
1.76	58.5891261744446\\
1.77	58.5820067395605\\
1.78	58.5749150834791\\
1.79	58.5678512154035\\
1.8	58.5608151109276\\
1.81	58.5538067263849\\
1.82	58.5468260166324\\
1.83	58.5398728725353\\
1.84	58.5329471725296\\
1.85	58.5260488209842\\
1.86	58.51917769273\\
1.87	58.5123336623593\\
1.88	58.5055166209017\\
1.89	58.4987263159282\\
1.9	58.491962523775\\
1.91	58.4852249974836\\
1.92	58.4785134364832\\
1.93	58.4718275642682\\
1.94	58.4651671150743\\
1.95	58.4585317725361\\
1.96	58.4519211848451\\
1.97	58.4453350007204\\
1.98	58.4387728121166\\
1.99	58.4322342219286\\
2	58.4257188905448\\
2.01	58.4192264280301\\
2.02	58.4127562640238\\
2.03	58.4063060687182\\
2.04	58.3998732845052\\
2.05	58.3934566219336\\
2.06	58.3870557994354\\
2.07	58.3806709979688\\
2.08	58.3743027219118\\
2.09	58.3679515589953\\
2.1	58.3616179991425\\
2.11	58.3553025469095\\
2.12	58.3490056998697\\
2.13	58.342727766617\\
2.14	58.3364689996039\\
2.15	58.330229607736\\
2.16	58.3240096932107\\
2.17	58.317809347868\\
2.18	58.3116286599528\\
2.19	58.3054676295508\\
2.2	58.2993262492946\\
2.21	58.2932045134821\\
2.22	58.2871023442822\\
2.23	58.28101967692\\
2.24	58.2749564691423\\
2.25	58.2689126184287\\
2.26	58.2628880363252\\
2.27	58.2568826603788\\
2.28	58.2508963790897\\
2.29	58.2449297320113\\
2.3	58.2389885071109\\
2.31	58.233079592254\\
2.32	58.227205705947\\
2.33	58.2213675502711\\
2.34	58.2155650762342\\
2.35	58.2097975007464\\
2.36	58.2040638249699\\
2.37	58.1983629658569\\
2.38	58.1926938459603\\
2.39	58.187055515355\\
2.4	58.1814470973601\\
2.41	58.1758678565386\\
2.42	58.1703171394815\\
2.43	58.1647942789677\\
2.44	58.1592987695613\\
2.45	58.1538302060767\\
2.46	58.1483881657625\\
2.47	58.1429723160499\\
2.48	58.1375824125695\\
2.49	58.1322182756077\\
2.5	58.1268798405134\\
2.51	58.1215670506009\\
2.52	58.1162797303154\\
2.53	58.1110176916496\\
2.54	58.105780750206\\
2.55	58.1005686658854\\
2.56	58.0953812314316\\
2.57	58.0902181621961\\
2.58	58.0850789976728\\
2.59	58.0799632929975\\
2.6	58.074870505233\\
2.61	58.0697998892553\\
2.62	58.0647506894876\\
2.63	58.0597220966956\\
2.64	58.0547132054857\\
2.65	58.049723187493\\
2.66	58.0447512027643\\
2.67	58.039796342653\\
2.68	58.0348577904982\\
2.69	58.0299347121579\\
2.7	58.0250262099766\\
2.71	58.0201315069636\\
2.72	58.0152498972799\\
2.73	58.010380686236\\
2.74	58.0055232580831\\
2.75	58.0006769004624\\
2.76	57.9958408833869\\
2.77	57.9910145080145\\
2.78	57.9861970080662\\
2.79	57.9813875404866\\
2.8	57.9765853178474\\
2.81	57.9717894627956\\
2.82	57.9669990289633\\
2.83	57.9622131382283\\
2.84	57.957430822385\\
2.85	57.9526510939793\\
2.86	57.9478735447009\\
2.87	57.9430984974513\\
2.88	57.9383265315729\\
2.89	57.9335583194219\\
2.9	57.9287944992206\\
2.91	57.924035698469\\
2.92	57.9192825661892\\
2.93	57.9145356134961\\
2.94	57.90979529963\\
2.95	57.9050621103851\\
2.96	57.9003363925399\\
2.97	57.8956185461066\\
2.98	57.890909009554\\
2.99	57.8862081155111\\
3	57.8815162105764\\
3.01	57.8768336729198\\
3.02	57.8721607798525\\
3.03	57.8674978578251\\
3.04	57.8628453403016\\
3.05	57.8582036534051\\
3.06	57.8535733139552\\
3.07	57.8489548746639\\
3.08	57.8443488574088\\
3.09	57.8397559484174\\
3.1	57.8351769293931\\
3.11	57.8306125750759\\
3.12	57.8260637877416\\
3.13	57.8215315121779\\
3.14	57.8170166437746\\
3.15	57.8125199359361\\
3.16	57.8080404714879\\
3.17	57.8035769465175\\
3.18	57.7991286965277\\
3.19	57.7946955500414\\
3.2	57.7902775895686\\
3.21	57.7858751580846\\
3.22	57.7814886403911\\
3.23	57.7771184274094\\
3.24	57.7727649950551\\
3.25	57.7684287608769\\
3.26	57.7641101115339\\
3.27	57.7598094954957\\
3.28	57.7555272974618\\
3.29	57.751263893173\\
3.3	57.7470197100872\\
3.31	57.7427951004938\\
3.32	57.7385904192837\\
3.33	57.7344060637328\\
3.34	57.7302423596087\\
3.35	57.7260997074264\\
3.36	57.7219786040124\\
3.37	57.717879506212\\
3.38	57.7138029115179\\
3.39	57.7097492829573\\
3.4	57.7057190174934\\
3.41	57.7017125003181\\
3.42	57.6977301147822\\
3.43	57.6937721712545\\
3.44	57.6898390381708\\
3.45	57.6859310885682\\
3.46	57.6820486410999\\
3.47	57.6781920877056\\
3.48	57.6743618131919\\
3.49	57.6705581580772\\
3.5	57.6667815373796\\
3.51	57.6630323327097\\
3.52	57.6593108731971\\
3.53	57.6556175391418\\
3.54	57.651952645262\\
3.55	57.6483164711755\\
3.56	57.6447093624349\\
3.57	57.6411316001362\\
3.58	57.637583386897\\
3.59	57.634065023906\\
3.6	57.6305768187988\\
3.61	57.6271191847159\\
3.62	57.6236925955027\\
3.63	57.6202976536991\\
3.64	57.6169360761469\\
3.65	57.6136102885411\\
3.66	57.6103225334035\\
3.67	57.6070749477288\\
3.68	57.6038696165633\\
3.69	57.6007084811822\\
3.7	57.5975934728339\\
3.71	57.5945264309126\\
3.72	57.5915090364942\\
3.73	57.588542998884\\
3.74	57.5856299906164\\
3.75	57.5827716384337\\
3.76	57.5799696340427\\
3.77	57.5772255757329\\
3.78	57.5745408296836\\
3.79	57.5719167359194\\
3.8	57.5693545660155\\
3.81	57.5668555683026\\
3.82	57.5644210386223\\
3.83	57.5620521897363\\
3.84	57.5597502287553\\
3.85	57.5575163822485\\
3.86	57.5553517791021\\
3.87	57.5532575783634\\
3.88	57.5512349465101\\
3.89	57.5492849648651\\
3.9	57.547408680233\\
3.91	57.5456067322806\\
3.92	57.5438793438554\\
3.93	57.5422268894581\\
3.94	57.5406497310355\\
3.95	57.539147405433\\
3.96	57.5377126412373\\
3.97	0\\
3.98	0\\
3.99	0\\
4	0\\
4.01	0\\
4.02	0\\
4.03	0\\
4.04	0\\
4.05	0\\
4.06	0\\
4.07	0\\
4.08	0\\
4.09	0\\
4.1	0\\
4.11	0\\
4.12	0\\
4.13	0\\
4.14	0\\
4.15	0\\
4.16	0\\
4.17	0\\
4.18	0\\
4.19	0\\
4.2	0\\
4.21	0\\
4.22	0\\
4.23	0\\
4.24	0\\
4.25	0\\
4.26	0\\
4.27	0\\
4.28	0\\
4.29	0\\
4.3	0\\
4.31	0\\
4.32	0\\
4.33	0\\
4.34	0\\
4.35	0\\
4.36	0\\
4.37	0\\
4.38	0\\
4.39	0\\
4.4	0\\
4.41	0\\
4.42	0\\
4.43	0\\
4.44	0\\
4.45	0\\
4.46	0\\
4.47	0\\
4.48	0\\
4.49	0\\
4.5	0\\
4.51	0\\
4.52	0\\
4.53	0\\
4.54	0\\
4.55	0\\
4.56	0\\
4.57	0\\
4.58	0\\
4.59	0\\
4.6	0\\
4.61	0\\
4.62	0\\
4.63	0\\
4.64	0\\
4.65	0\\
4.66	0\\
4.67	0\\
4.68	0\\
4.69	0\\
4.7	0\\
4.71	0\\
4.72	0\\
4.73	0\\
4.74	0\\
4.75	0\\
4.76	0\\
4.77	0\\
4.78	0\\
4.79	0\\
4.8	0\\
4.81	0\\
4.82	0\\
4.83	0\\
4.84	0\\
4.85	0\\
4.86	0\\
4.87	0\\
4.88	0\\
4.89	0\\
4.9	0\\
4.91	0\\
4.92	0\\
4.93	0\\
4.94	0\\
4.95	0\\
4.96	0\\
4.97	0\\
4.98	0\\
4.99	0\\
5	0\\
5.01	0\\
5.02	0\\
5.03	0\\
5.04	0\\
5.05	0\\
5.06	0\\
5.07	0\\
5.08	0\\
5.09	0\\
5.1	0\\
5.11	0\\
5.12	0\\
5.13	0\\
5.14	0\\
5.15	0\\
5.16	0\\
5.17	0\\
5.18	0\\
5.19	0\\
5.2	0\\
5.21	0\\
5.22	0\\
5.23	0\\
5.24	0\\
5.25	0\\
5.26	0\\
5.27	0\\
5.28	0\\
5.29	0\\
5.3	0\\
5.31	0\\
5.32	0\\
5.33	0\\
5.34	0\\
5.35	0\\
5.36	0\\
5.37	0\\
5.38	0\\
5.39	0\\
5.4	0\\
5.41	0\\
5.42	0\\
5.43	0\\
5.44	0\\
5.45	0\\
5.46	0\\
5.47	0\\
5.48	0\\
5.49	0\\
5.5	0\\
5.51	0\\
5.52	0\\
5.53	0\\
5.54	0\\
5.55	0\\
5.56	0\\
5.57	0\\
5.58	0\\
5.59	0\\
5.6	0\\
5.61	0\\
5.62	0\\
5.63	0\\
5.64	0\\
5.65	0\\
5.66	0\\
5.67	0\\
5.68	0\\
5.69	0\\
5.7	0\\
5.71	0\\
5.72	0\\
5.73	0\\
5.74	0\\
5.75	0\\
5.76	0\\
5.77	0\\
5.78	0\\
5.79	0\\
5.8	0\\
5.81	0\\
5.82	0\\
5.83	0\\
5.84	0\\
5.85	0\\
5.86	0\\
5.87	0\\
5.88	0\\
5.89	0\\
5.9	0\\
5.91	0\\
5.92	0\\
5.93	0\\
5.94	0\\
5.95	0\\
5.96	0\\
5.97	0\\
5.98	0\\
5.99	0\\
6	0\\
6.01	0\\
6.02	0\\
6.03	0\\
6.04	0\\
6.05	0\\
6.06	0\\
6.07	0\\
6.08	0\\
6.09	0\\
6.1	0\\
6.11	0\\
6.12	0\\
6.13	0\\
6.14	0\\
6.15	0\\
6.16	0\\
6.17	0\\
6.18	0\\
6.19	0\\
6.2	0\\
6.21	0\\
6.22	0\\
6.23	0\\
6.24	0\\
6.25	0\\
6.26	0\\
6.27	0\\
6.28	0\\
6.29	0\\
6.3	0\\
6.31	0\\
6.32	0\\
6.33	0\\
6.34	0\\
6.35	0\\
6.36	0\\
6.37	0\\
6.38	0\\
6.39	0\\
6.4	0\\
6.41	0\\
6.42	0\\
6.43	0\\
6.44	0\\
6.45	0\\
6.46	0\\
6.47	0\\
6.48	0\\
6.49	0\\
6.5	0\\
6.51	0\\
6.52	0\\
6.53	0\\
6.54	0\\
6.55	0\\
6.56	0\\
6.57	0\\
6.58	0\\
6.59	0\\
6.6	0\\
6.61	0\\
6.62	0\\
6.63	0\\
6.64	0\\
6.65	0\\
6.66	0\\
6.67	0\\
6.68	0\\
6.69	0\\
6.7	0\\
6.71	0\\
6.72	0\\
6.73	0\\
6.74	0\\
6.75	0\\
6.76	0\\
6.77	0\\
6.78	0\\
6.79	0\\
6.8	0\\
6.81	0\\
6.82	0\\
6.83	0\\
6.84	0\\
6.85	0\\
6.86	0\\
6.87	0\\
6.88	0\\
6.89	0\\
6.9	0\\
6.91	0\\
6.92	0\\
6.93	0\\
6.94	0\\
6.95	0\\
6.96	0\\
6.97	0\\
6.98	0\\
6.99	0\\
7	0\\
7.01	0\\
7.02	0\\
7.03	0\\
7.04	0\\
7.05	0\\
7.06	0\\
7.07	0\\
7.08	0\\
7.09	0\\
7.1	0\\
7.11	0\\
7.12	0\\
7.13	0\\
7.14	0\\
7.15	0\\
7.16	0\\
7.17	0\\
7.18	0\\
7.19	0\\
7.2	0\\
7.21	0\\
7.22	0\\
7.23	0\\
7.24	0\\
7.25	0\\
7.26	0\\
7.27	0\\
7.28	0\\
7.29	0\\
7.3	0\\
7.31	0\\
7.32	0\\
7.33	0\\
7.34	0\\
7.35	0\\
7.36	0\\
7.37	0\\
7.38	0\\
7.39	0\\
7.4	0\\
7.41	0\\
7.42	0\\
7.43	0\\
7.44	0\\
7.45	0\\
7.46	0\\
7.47	0\\
7.48	0\\
7.49	0\\
7.5	0\\
7.51	0\\
7.52	0\\
7.53	0\\
7.54	0\\
7.55	0\\
7.56	0\\
7.57	0\\
7.58	0\\
7.59	0\\
7.6	0\\
7.61	0\\
7.62	0\\
7.63	0\\
7.64	0\\
7.65	0\\
7.66	0\\
7.67	0\\
7.68	0\\
7.69	0\\
7.7	0\\
7.71	0\\
7.72	0\\
7.73	0\\
7.74	0\\
7.75	0\\
7.76	0\\
7.77	0\\
7.78	0\\
7.79	0\\
7.8	0\\
7.81	0\\
7.82	0\\
7.83	0\\
7.84	0\\
7.85	0\\
7.86	0\\
7.87	0\\
7.88	0\\
7.89	0\\
7.9	0\\
7.91	0\\
7.92	0\\
7.93	0\\
7.94	0\\
7.95	0\\
7.96	0\\
7.97	0\\
7.98	0\\
7.99	0\\
8	0\\
8.01	0\\
8.02	0\\
8.03	0\\
8.04	0\\
8.05	0\\
8.06	0\\
8.07	0\\
8.08	0\\
8.09	0\\
8.1	0\\
8.11	0\\
8.12	0\\
8.13	0\\
8.14	0\\
8.15	0\\
8.16	0\\
8.17	0\\
8.18	0\\
8.19	0\\
8.2	0\\
8.21	0\\
8.22	0\\
8.23	0\\
8.24	0\\
8.25	0\\
8.26	0\\
8.27	0\\
8.28	0\\
8.29	0\\
8.3	0\\
8.31	0\\
8.32	0\\
8.33	0\\
8.34	0\\
8.35	0\\
8.36	0\\
8.37	0\\
8.38	0\\
8.39	0\\
8.4	0\\
8.41	0\\
8.42	0\\
8.43	0\\
8.44	0\\
8.45	0\\
8.46	0\\
8.47	0\\
8.48	0\\
8.49	0\\
8.5	0\\
8.51	0\\
8.52	0\\
8.53	0\\
8.54	0\\
8.55	0\\
8.56	0\\
8.57	0\\
8.58	0\\
8.59	0\\
8.6	0\\
8.61	0\\
8.62	0\\
8.63	0\\
8.64	0\\
8.65	0\\
8.66	0\\
8.67	0\\
8.68	0\\
8.69	0\\
8.7	0\\
8.71	0\\
8.72	0\\
8.73	0\\
8.74	0\\
8.75	0\\
8.76	0\\
8.77	0\\
8.78	0\\
8.79	0\\
8.8	0\\
8.81	0\\
8.82	0\\
8.83	0\\
8.84	0\\
8.85	0\\
8.86	0\\
8.87	0\\
8.88	0\\
8.89	0\\
8.9	0\\
8.91	0\\
8.92	0\\
8.93	0\\
8.94	0\\
8.95	0\\
8.96	0\\
8.97	0\\
8.98	0\\
8.99	0\\
9	0\\
9.01	0\\
9.02	0\\
9.03	0\\
9.04	0\\
9.05	0\\
9.06	0\\
9.07	0\\
9.08	0\\
9.09	0\\
9.1	0\\
9.11	0\\
9.12	0\\
9.13	0\\
9.14	0\\
9.15	0\\
9.16	0\\
9.17	0\\
9.18	0\\
9.19	0\\
9.2	0\\
9.21	0\\
9.22	0\\
9.23	0\\
9.24	0\\
9.25	0\\
9.26	0\\
9.27	0\\
9.28	0\\
9.29	0\\
9.3	0\\
9.31	0\\
9.32	0\\
9.33	0\\
9.34	0\\
9.35	0\\
9.36	0\\
9.37	0\\
9.38	0\\
9.39	0\\
9.4	0\\
9.41	0\\
9.42	0\\
9.43	0\\
9.44	0\\
9.45	0\\
9.46	0\\
9.47	0\\
9.48	0\\
9.49	0\\
9.5	0\\
9.51	0\\
9.52	0\\
9.53	0\\
9.54	0\\
9.55	0\\
9.56	0\\
9.57	0\\
9.58	0\\
9.59	0\\
9.6	0\\
9.61	0\\
9.62	0\\
9.63	0\\
9.64	0\\
9.65	0\\
9.66	0\\
9.67	0\\
9.68	0\\
9.69	0\\
9.7	0\\
9.71	0\\
9.72	0\\
9.73	0\\
9.74	0\\
9.75	0\\
9.76	0\\
9.77	0\\
9.78	0\\
9.79	0\\
9.8	0\\
9.81	0\\
9.82	0\\
9.83	0\\
9.84	0\\
9.85	0\\
9.86	0\\
9.87	0\\
9.88	0\\
9.89	0\\
9.9	0\\
9.91	0\\
9.92	0\\
9.93	0\\
9.94	0\\
9.95	0\\
9.96	0\\
9.97	0\\
9.98	0\\
9.99	0\\
10	0\\
10.01	0\\
10.02	0\\
10.03	0\\
10.04	0\\
10.05	0\\
10.06	0\\
10.07	0\\
10.08	0\\
10.09	0\\
10.1	0\\
10.11	0\\
10.12	0\\
10.13	0\\
10.14	0\\
10.15	0\\
10.16	0\\
10.17	0\\
10.18	0\\
10.19	0\\
10.2	0\\
10.21	0\\
10.22	0\\
10.23	0\\
10.24	0\\
10.25	0\\
10.26	0\\
10.27	0\\
10.28	0\\
10.29	0\\
10.3	0\\
10.31	0\\
10.32	0\\
10.33	0\\
10.34	0\\
10.35	0\\
10.36	0\\
10.37	0\\
10.38	0\\
10.39	0\\
10.4	0\\
10.41	0\\
10.42	0\\
10.43	0\\
10.44	0\\
10.45	0\\
10.46	0\\
10.47	0\\
10.48	0\\
10.49	0\\
10.5	0\\
10.51	0\\
10.52	0\\
10.53	0\\
10.54	0\\
10.55	0\\
10.56	0\\
10.57	0\\
10.58	0\\
10.59	0\\
10.6	0\\
10.61	0\\
10.62	0\\
10.63	0\\
10.64	0\\
10.65	0\\
10.66	0\\
10.67	0\\
10.68	0\\
10.69	0\\
10.7	0\\
10.71	0\\
10.72	0\\
10.73	0\\
10.74	0\\
10.75	0\\
10.76	0\\
10.77	0\\
10.78	0\\
10.79	0\\
10.8	0\\
10.81	0\\
10.82	0\\
10.83	0\\
10.84	0\\
10.85	0\\
10.86	0\\
10.87	0\\
10.88	0\\
10.89	0\\
10.9	0\\
10.91	0\\
10.92	0\\
10.93	0\\
10.94	0\\
10.95	0\\
10.96	0\\
10.97	0\\
10.98	0\\
10.99	0\\
11	0\\
};
\addlegendentry{Behavioral Cloning}

\addplot [color=black, line width=1.0pt]
  table[row sep=crcr]{%
0	60.2559940743791\\
0.01	60.2446184290356\\
0.02	60.2329004407448\\
0.03	60.2212231567379\\
0.04	60.2096211830429\\
0.05	60.1980194751162\\
0.06	60.1864475463038\\
0.07	60.1749068497201\\
0.08	60.1633872004\\
0.09	60.1518931287715\\
0.1	60.1404248551005\\
0.11	60.1289823328926\\
0.12	60.1175666563038\\
0.13	60.1061778125067\\
0.14	60.0948156442701\\
0.15	60.0834802022098\\
0.16	60.0721713598853\\
0.17	60.0608889883543\\
0.18	60.0496331104533\\
0.19	60.0384037793492\\
0.2	60.0272008651921\\
0.21	60.0160242116242\\
0.22	60.0048736521229\\
0.23	59.9937490269279\\
0.24	59.9826501842002\\
0.25	59.9715769765031\\
0.26	59.9605292566344\\
0.27	59.9495068711307\\
0.28	59.9385096543852\\
0.29	59.9275373491366\\
0.3	59.9165896329977\\
0.31	59.9056662349065\\
0.32	59.8947668828194\\
0.33	59.8838913179022\\
0.34	59.8730393064567\\
0.35	59.8622106183358\\
0.36	59.8514050601501\\
0.37	59.8406224934944\\
0.38	59.8298628317048\\
0.39	59.819126030085\\
0.4	59.8084120755752\\
0.41	59.7977209777592\\
0.42	59.7870527627278\\
0.43	59.7764074690094\\
0.44	59.7657851441116\\
0.45	59.75518584291\\
0.46	59.7446096263277\\
0.47	59.7340565593571\\
0.48	59.7235267109426\\
0.49	59.713020151153\\
0.5	59.7025369512512\\
0.51	59.6920771820241\\
0.52	59.681640913095\\
0.53	59.6712282121068\\
0.54	59.6608391441373\\
0.55	59.6504737711878\\
0.56	59.6401321517886\\
0.57	59.6298143406864\\
0.58	59.6195203887011\\
0.59	59.6092503424222\\
0.6	59.5990042443244\\
0.61	59.5887821326717\\
0.62	59.5785840415437\\
0.63	59.5684100009356\\
0.64	59.5582600368855\\
0.65	59.5481341716025\\
0.66	59.5380324236464\\
0.67	59.5279548081434\\
0.68	59.517901336964\\
0.69	59.5078720189632\\
0.7	59.497866860212\\
0.71	59.4878858642148\\
0.72	59.4779290321504\\
0.73	59.4679963631136\\
0.74	59.4580878543211\\
0.75	59.4482035013463\\
0.76	59.4383432983415\\
0.77	59.4285072382248\\
0.78	59.4186953128858\\
0.79	59.4089075133983\\
0.8	59.3991438323358\\
0.81	59.3894042640461\\
0.82	59.3796887994156\\
0.83	59.3699974284224\\
0.84	59.3603301420281\\
0.85	59.3506869326985\\
0.86	59.3410677938086\\
0.87	59.331472719019\\
0.88	59.3219017014942\\
0.89	59.3123547335869\\
0.9	59.3028318069185\\
0.91	59.2933329126376\\
0.92	59.2838580418048\\
0.93	59.2744071856688\\
0.94	59.2649803357862\\
0.95	59.2555774840118\\
0.96	59.2461986224293\\
0.97	59.2368437432749\\
0.98	59.2275128389145\\
0.99	59.2182059018784\\
1	59.2089229249226\\
1.01	59.1996639011094\\
1.02	59.1904288238777\\
1.03	59.1812176628314\\
1.04	59.172030431616\\
1.05	59.1628671334627\\
1.06	59.1537277635038\\
1.07	59.1446123171488\\
1.08	59.1355207900075\\
1.09	59.1264531782522\\
1.1	59.1174094784929\\
1.11	59.1083896877567\\
1.12	59.0993938034935\\
1.13	59.0904218236544\\
1.14	59.0814737467184\\
1.15	59.0725495717186\\
1.16	59.0636492981353\\
1.17	59.0547729256354\\
1.18	59.0459204545565\\
1.19	59.0370918863854\\
1.2	59.028287224057\\
1.21	59.0195064707159\\
1.22	59.0107496295156\\
1.23	59.0020167045136\\
1.24	58.993307684468\\
1.25	58.9846225382952\\
1.26	58.9759613251802\\
1.27	58.967324051236\\
1.28	58.9587107223456\\
1.29	58.9501213441994\\
1.3	58.9415559221378\\
1.31	58.9330144611704\\
1.32	58.9244969659144\\
1.33	58.9160034406676\\
1.34	58.9075338893872\\
1.35	58.89908831569\\
1.36	58.8906667228417\\
1.37	58.8822691137798\\
1.38	58.8738954911198\\
1.39	58.8655458571533\\
1.4	58.8572202138565\\
1.41	58.8489185628996\\
1.42	58.840640905669\\
1.43	58.832387243269\\
1.44	58.8241575765397\\
1.45	58.8159519060784\\
1.46	58.8077701838575\\
1.47	58.7996124411988\\
1.48	58.7914786952359\\
1.49	58.7833689455987\\
1.5	58.7752831917467\\
1.51	58.7672214330196\\
1.52	58.7591836686413\\
1.53	58.7511698977101\\
1.54	58.743180119248\\
1.55	58.7352143322222\\
1.56	58.7272725355742\\
1.57	58.7193547282464\\
1.58	58.7114609090422\\
1.59	58.7035910765167\\
1.6	58.6957452291998\\
1.61	58.6879233659767\\
1.62	58.6801254859192\\
1.63	58.6723515876857\\
1.64	58.6646016697918\\
1.65	58.6568757308779\\
1.66	58.6491737697585\\
1.67	58.6414957693898\\
1.68	58.6338416990842\\
1.69	58.6262116037177\\
1.7	58.6186054823428\\
1.71	58.6110233340682\\
1.72	58.6034651580395\\
1.73	58.5959309534798\\
1.74	58.5884207198258\\
1.75	58.5809344566591\\
1.76	58.5734721636879\\
1.77	58.5660338408352\\
1.78	58.5586194881121\\
1.79	58.5512291055887\\
1.8	58.5438626934853\\
1.81	58.5365202520357\\
1.82	58.5292017814783\\
1.83	58.5219072821436\\
1.84	58.5146367544765\\
1.85	58.5073901991365\\
1.86	58.5001676170374\\
1.87	58.4929690086767\\
1.88	58.4857943736909\\
1.89	58.4786436650067\\
1.9	58.4715169162743\\
1.91	58.4644141380547\\
1.92	58.45733532933\\
1.93	58.4502804890629\\
1.94	58.4432496162949\\
1.95	58.436242709298\\
1.96	58.4292597667931\\
1.97	58.4223007809197\\
1.98	58.4153657443751\\
1.99	58.4084546523793\\
2	58.4015675004858\\
2.01	58.3947042846459\\
2.02	58.3878650018183\\
2.03	58.3810496495555\\
2.04	58.3742582263947\\
2.05	58.3674907319054\\
2.06	58.3607471663857\\
2.07	58.3540275313651\\
2.08	58.3473318292487\\
2.09	58.3406600620158\\
2.1	58.3340122131952\\
2.11	58.3273882714149\\
2.12	58.3207882760921\\
2.13	58.314212232459\\
2.14	58.3076601456964\\
2.15	58.3011320204167\\
2.16	58.2946278609679\\
2.17	58.2881476709156\\
2.18	58.2816914529709\\
2.19	58.2752592094749\\
2.2	58.2688509419759\\
2.21	58.2624666512815\\
2.22	58.256106337917\\
2.23	58.2497700016804\\
2.24	58.2434576417497\\
2.25	58.2371692568782\\
2.26	58.2309048432701\\
2.27	58.2246643979515\\
2.28	58.2184479177242\\
2.29	58.2122553986209\\
2.3	58.2060868366898\\
2.31	58.1999422281954\\
2.32	58.1938215231265\\
2.33	58.1877247628199\\
2.34	58.1816519487893\\
2.35	58.1756030796881\\
2.36	58.1695781545244\\
2.37	58.1635771724084\\
2.38	58.1576001326775\\
2.39	58.1516470348756\\
2.4	58.1457178786149\\
2.41	58.1398126636904\\
2.42	58.1339313900427\\
2.43	58.1280740576344\\
2.44	58.1222406657904\\
2.45	58.116431215005\\
2.46	58.1106457058018\\
2.47	58.104884138755\\
2.48	58.0991465142959\\
2.49	58.0934328314219\\
2.5	58.087743087987\\
2.51	58.0820772808003\\
2.52	58.0764354064294\\
2.53	58.0708174375915\\
2.54	58.0652233750984\\
2.55	58.0596532405533\\
2.56	58.0541070349198\\
2.57	58.0485847604878\\
2.58	58.0430864203435\\
2.59	58.0376120180068\\
2.6	58.0321615575056\\
2.61	58.026735043212\\
2.62	58.0213324793801\\
2.63	58.0159538705363\\
2.64	58.0105992174905\\
2.65	58.0052685184146\\
2.66	57.999961770382\\
2.67	57.9946789699391\\
2.68	57.9894201137237\\
2.69	57.9841851986576\\
2.7	57.978974222175\\
2.71	57.9737871824052\\
2.72	57.9686240779948\\
2.73	57.9634849080091\\
2.74	57.9583696660485\\
2.75	57.953278321443\\
2.76	57.9482109111454\\
2.77	57.9431674358743\\
2.78	57.9381478965122\\
2.79	57.9331522940351\\
2.8	57.9281806295845\\
2.81	57.9232329042358\\
2.82	57.918309119025\\
2.83	57.9134092751012\\
2.84	57.9085333734239\\
2.85	57.9036814147184\\
2.86	57.8988533988514\\
2.87	57.8940493256432\\
2.88	57.8892691954295\\
2.89	57.8845130088518\\
2.9	57.87978076661\\
2.91	57.8750724694461\\
2.92	57.8703881180776\\
2.93	57.8657277129125\\
2.94	57.8610912541704\\
2.95	57.8564787419652\\
2.96	57.8518901461531\\
2.97	57.8473254875051\\
2.98	57.8427847744576\\
2.99	57.8382680062199\\
3	57.8337751804817\\
3.01	57.8293062955173\\
3.02	57.8248613491481\\
3.03	57.8204403386312\\
3.04	57.8160432606286\\
3.05	57.8116701117819\\
3.06	57.807320888995\\
3.07	57.8029955896753\\
3.08	57.798694211891\\
3.09	57.7944167540941\\
3.1	57.7901632151124\\
3.11	57.7859335941917\\
3.12	57.7817278907765\\
3.13	57.7775461045431\\
3.14	57.7733882354486\\
3.15	57.7692542835745\\
3.16	57.7651442491693\\
3.17	57.7610581179111\\
3.18	57.7569958846743\\
3.19	57.7529575702009\\
3.2	57.7489431750065\\
3.21	57.7449526995747\\
3.22	57.7409861444545\\
3.23	57.7370435101785\\
3.24	57.7331247972296\\
3.25	57.7292300061648\\
3.26	57.72535913748\\
3.27	57.7215121915963\\
3.28	57.717689169007\\
3.29	57.7138900701112\\
3.3	57.7101148952122\\
3.31	57.7063636446999\\
3.32	57.7026363188533\\
3.33	57.6989329178532\\
3.34	57.6952534419997\\
3.35	57.6915978914237\\
3.36	57.6879662659063\\
3.37	57.6843585651849\\
3.38	57.6807747863026\\
3.39	57.6772149034661\\
3.4	57.6736789450465\\
3.41	57.6701669109888\\
3.42	57.6666788011299\\
3.43	57.6632146153668\\
3.44	57.6597743534676\\
3.45	57.6563580150746\\
3.46	57.6529655999043\\
3.47	57.6495971075901\\
3.48	57.6462525376866\\
3.49	57.6429318898309\\
3.5	57.6396351636073\\
3.51	57.6363623585505\\
3.52	57.6331134742898\\
3.53	57.6298885104536\\
3.54	57.6266874666864\\
3.55	57.6235103425151\\
3.56	57.6203571375057\\
3.57	57.617227852021\\
3.58	57.614122485525\\
3.59	57.6110410370542\\
3.6	57.6079834847519\\
3.61	57.6049498428482\\
3.62	57.6019401170159\\
3.63	57.5989543067855\\
3.64	57.5959924117423\\
3.65	57.5930544315892\\
3.66	57.5901403661024\\
3.67	57.5872502152843\\
3.68	57.5843839794671\\
3.69	57.5815416587812\\
3.7	57.5787232530358\\
3.71	57.575928762548\\
3.72	57.5731581879594\\
3.73	57.5704115302901\\
3.74	57.5676887910225\\
3.75	57.5649899717111\\
3.76	57.5623150741152\\
3.77	57.5596641002469\\
3.78	57.5570370520101\\
3.79	57.5544339314444\\
3.8	57.551854740778\\
3.81	57.5492994709023\\
3.82	57.5467681227592\\
3.83	57.5442607106618\\
3.84	57.5417772372882\\
3.85	57.5393177081599\\
3.86	57.5368821311516\\
3.87	57.5344705111001\\
3.88	57.5320828484902\\
3.89	57.5297191415121\\
3.9	57.5273793868321\\
3.91	57.5250635807596\\
3.92	57.5227717197849\\
3.93	57.5205038004012\\
3.94	57.5182598195858\\
3.95	57.5160397749373\\
3.96	57.5138436620678\\
3.97	57.5116714604867\\
3.98	57.5095231464449\\
3.99	57.5073987054963\\
4	57.5052981241143\\
4.01	57.5032213903431\\
4.02	57.5011684915503\\
4.03	57.4991394088286\\
4.04	57.4971341538779\\
4.05	57.4951527257406\\
4.06	57.4931951256938\\
4.07	57.4912613586194\\
4.08	57.4893514316344\\
4.09	57.4874653511615\\
4.1	57.4856031263499\\
4.11	57.4837647696095\\
4.12	57.4819502948311\\
4.13	57.4801597177493\\
4.14	57.4783930537614\\
4.15	57.4766503168479\\
4.16	57.4749315202872\\
4.17	57.4732366750058\\
4.18	57.4715657896633\\
4.19	57.4699188718368\\
4.2	57.4682959266984\\
4.21	57.4666969573571\\
4.22	57.4651219659273\\
4.23	57.4635709522433\\
4.24	57.4620438992216\\
4.25	57.4605408108607\\
4.26	57.4590616808105\\
4.27	57.4576065007318\\
4.28	57.4561752587035\\
4.29	57.4547679429584\\
4.3	57.4533845430635\\
4.31	57.4520250499515\\
4.32	57.4506894566485\\
4.33	57.4493777578861\\
4.34	57.4480899491802\\
4.35	57.4468260270611\\
4.36	57.4455859888946\\
4.37	57.4443698323636\\
4.38	57.443177555849\\
4.39	57.4420091584002\\
4.4	57.440864639329\\
4.41	57.439743998435\\
4.42	57.4386472359008\\
4.43	57.4375743519902\\
4.44	57.4365253458515\\
4.45	57.4355002117084\\
4.46	57.4344989581363\\
4.47	57.433521592466\\
4.48	57.4325681179274\\
4.49	57.4316385361202\\
4.5	57.4307328469342\\
4.51	57.4298510470074\\
4.52	57.4289931327005\\
4.53	57.42815910005\\
4.54	57.4273489445252\\
4.55	57.4265626628367\\
4.56	57.4258002536619\\
4.57	57.4250617180633\\
4.58	57.4243470587266\\
4.59	57.4236562796616\\
4.6	57.4229893863963\\
4.61	57.4223463865759\\
4.62	57.4217272845638\\
4.63	57.4211320835789\\
4.64	57.4205607834002\\
4.65	57.420013381447\\
4.66	57.4194898700475\\
4.67	57.4189902442386\\
4.68	57.4185145041169\\
4.69	57.4180626462393\\
4.7	57.4176346678858\\
4.71	57.4172305671375\\
4.72	57.4168503427569\\
4.73	57.4164939940246\\
4.74	57.4161615209839\\
4.75	57.415852924232\\
4.76	57.4155682046847\\
4.77	57.4153073637383\\
4.78	57.4150704029755\\
4.79	57.4148573238809\\
4.8	57.4146681280965\\
4.81	57.4145028172471\\
4.82	57.4143613927684\\
4.83	57.4142438562617\\
4.84	57.4141502092981\\
4.85	57.4140804530271\\
4.86	57.4140345875586\\
4.87	57.4140126113835\\
4.88	57.4140145227986\\
4.89	57.4140403257255\\
4.9	57.4140900216641\\
4.91	57.4141636122884\\
4.92	57.4142610994688\\
4.93	57.4143824848882\\
4.94	57.4145277698362\\
4.95	57.4146969554659\\
4.96	57.4148900425865\\
4.97	57.4151070315819\\
4.98	57.415347922641\\
4.99	57.4156127155238\\
5	57.415901408456\\
5.01	57.4162043917044\\
5.02	57.4165120560522\\
5.03	57.4168243954839\\
5.04	57.4171414033\\
5.05	57.4174630733349\\
5.06	57.4177894002205\\
5.07	57.4181203793404\\
5.08	57.418456007099\\
5.09	57.4187962807004\\
5.1	57.4191411979501\\
5.11	57.4194907574369\\
5.12	57.4198449583538\\
5.13	57.4202038003748\\
5.14	57.4205672837813\\
5.15	57.4209354092079\\
5.16	57.4213081775091\\
5.17	57.4216855898163\\
5.18	57.4220676473014\\
5.19	57.4224543511657\\
5.2	57.4228457027217\\
5.21	57.4232417032069\\
5.22	57.4236423538388\\
5.23	57.4240476558953\\
5.24	57.4244576104801\\
5.25	57.4248722186237\\
5.26	57.4252914813733\\
5.27	57.4257153995473\\
5.28	57.4261439738888\\
5.29	57.4265772051582\\
5.3	57.4270150972892\\
5.31	57.427457647833\\
5.32	57.4279048567997\\
5.33	57.4283567244764\\
5.34	57.4288132508098\\
5.35	57.4292744355449\\
5.36	57.429740278413\\
5.37	57.4302107792894\\
5.38	57.4306859382357\\
5.39	57.4311657551955\\
5.4	57.4316502300709\\
5.41	57.4321393628356\\
5.42	57.4326331532388\\
5.43	57.433131600937\\
5.44	57.4336347056699\\
5.45	57.4341424670164\\
5.46	57.4346548845041\\
5.47	57.4351719577655\\
5.48	57.4356936863135\\
5.49	57.4362200696104\\
5.5	57.436751107203\\
5.51	57.4372868037509\\
5.52	57.4378271560961\\
5.53	57.4383721610945\\
5.54	57.4389218176944\\
5.55	57.4394761253045\\
5.56	57.4400350833429\\
5.57	57.440598691183\\
5.58	57.4411669488649\\
5.59	57.4417398555826\\
5.6	57.4423174102624\\
5.61	57.4428996121709\\
5.62	57.4434864608286\\
5.63	57.4440779559215\\
5.64	57.4446740971351\\
5.65	57.4452748841604\\
5.66	57.4458803167509\\
5.67	57.4464903946967\\
5.68	57.4471051180067\\
5.69	57.4477244870125\\
5.7	57.4483485021649\\
5.71	57.4489771640425\\
5.72	57.4496104780623\\
5.73	57.4502484452434\\
5.74	57.45089105954\\
5.75	57.4515383218789\\
5.76	57.4521902333847\\
5.77	57.4528467955593\\
5.78	57.4535080102587\\
5.79	57.4541738792686\\
5.8	57.4548444044867\\
5.81	57.4555195880114\\
5.82	57.4561994317462\\
5.83	57.456883937606\\
5.84	57.4575731076008\\
5.85	57.4582669439444\\
5.86	57.4589654503467\\
5.87	57.4596686316741\\
5.88	57.4603764907828\\
5.89	57.4610890278997\\
5.9	57.46180624191\\
5.91	57.4625281306923\\
5.92	57.463254691832\\
5.93	57.4639859265174\\
5.94	57.4647218409103\\
5.95	57.4654624206086\\
5.96	57.4662076629643\\
5.97	57.466957564288\\
5.98	57.4677121001747\\
5.99	57.4684712482218\\
6	57.4692349956263\\
6.01	57.4700033310081\\
6.02	57.4707762445993\\
6.03	57.4715537298173\\
6.04	57.472335781745\\
6.05	57.473122398155\\
6.06	57.4739135797343\\
6.07	57.474709328913\\
6.08	57.4755096505538\\
6.09	57.4763145498051\\
6.1	57.4771240331264\\
6.11	57.4779381112027\\
6.12	57.4787567972646\\
6.13	57.4795801051734\\
6.14	57.4804080498213\\
6.15	57.481240664875\\
6.16	57.4820779436355\\
6.17	57.4829198973666\\
6.18	57.4837665358424\\
6.19	57.4846178663937\\
6.2	57.4854738953068\\
6.21	57.4863346271268\\
6.22	57.4872000639999\\
6.23	57.488070207057\\
6.24	57.4889450556513\\
6.25	57.4898246028765\\
6.26	57.4907088401936\\
6.27	57.4915977606433\\
6.28	57.4924913538157\\
6.29	57.493389608149\\
6.3	57.4942925127918\\
6.31	57.4952000588018\\
6.32	57.4961122388459\\
6.33	57.4970290470812\\
6.34	57.4979504792031\\
6.35	57.4988765316154\\
6.36	57.4998072187171\\
6.37	57.5007425269554\\
6.38	57.501682448643\\
6.39	57.5026269825327\\
6.4	57.5035761280742\\
6.41	57.5045298850197\\
6.42	57.5054882533526\\
6.43	57.5064512334214\\
6.44	57.5074188256437\\
6.45	57.5083910305243\\
6.46	57.5093678473686\\
6.47	57.510349278681\\
6.48	57.5113353284872\\
6.49	57.5123260011087\\
6.5	57.5133213008578\\
6.51	57.5143212301844\\
6.52	57.5153257897104\\
6.53	57.5163349765351\\
6.54	57.5173487865764\\
6.55	57.5183672157681\\
6.56	57.5193902587841\\
6.57	57.5204179245693\\
6.58	57.5214502111082\\
6.59	57.5224871041769\\
6.6	57.5235286050345\\
6.61	57.5245747166095\\
6.62	57.5256254435228\\
6.63	57.5266807909992\\
6.64	57.5277407648409\\
6.65	57.5288053709999\\
6.66	57.5298746120269\\
6.67	57.5309484871397\\
6.68	57.5320269933088\\
6.69	57.5331101263221\\
6.7	57.534197881732\\
6.71	57.5352902552415\\
6.72	57.5363872429299\\
6.73	57.5374888416176\\
6.74	57.5385950488599\\
6.75	57.5397058627501\\
6.76	57.5408212821509\\
6.77	57.5419413064996\\
6.78	57.5430659427283\\
6.79	57.5441952086188\\
6.8	57.5453290809505\\
6.81	57.5464675608407\\
6.82	57.5476106497501\\
6.83	57.5487583492587\\
6.84	57.5499106609139\\
6.85	57.5510675865467\\
6.86	57.5522291280604\\
6.87	57.5533952872194\\
6.88	57.5545660655071\\
6.89	57.5557414630844\\
6.9	57.5569214801517\\
6.91	57.5581061179336\\
6.92	57.5592953781086\\
6.93	57.5604892626114\\
6.94	57.5616877737175\\
6.95	57.5628909135584\\
6.96	57.5640986839008\\
6.97	57.5653110863759\\
6.98	57.566528122249\\
6.99	57.5677497923531\\
7	57.5689761306336\\
7.01	57.5702071057387\\
7.02	57.5714427155967\\
7.03	57.5726829578063\\
7.04	57.5739278300397\\
7.05	57.5751773287477\\
7.06	57.5764314475556\\
7.07	57.5776901799838\\
7.08	57.57895352005\\
7.09	57.5802214623757\\
7.1	57.5814940026446\\
7.11	57.5827711374998\\
7.12	57.5840528642381\\
7.13	57.5853391809941\\
7.14	57.5866300865867\\
7.15	57.5879255802875\\
7.16	57.5892256619897\\
7.17	57.590530332045\\
7.18	57.5918395910356\\
7.19	57.5931534398631\\
7.2	57.5944718795755\\
7.21	57.5957949371394\\
7.22	57.597122600823\\
7.23	57.5984548587454\\
7.24	57.599791711985\\
7.25	57.6011331616843\\
7.26	57.6024792088734\\
7.27	57.6038298544954\\
7.28	57.6051850995061\\
7.29	57.6065449446986\\
7.3	57.6079093907376\\
7.31	57.609278438313\\
7.32	57.610652087938\\
7.33	57.6120303399971\\
7.34	57.6134131949379\\
7.35	57.6148006530344\\
7.36	57.6161927144343\\
7.37	57.6175893793848\\
7.38	57.6189906479309\\
7.39	57.6203965196637\\
7.4	57.6218069941644\\
7.41	57.6232220711152\\
7.42	57.6246417661086\\
7.43	57.6260660900709\\
7.44	57.6274950159593\\
7.45	57.628928543514\\
7.46	57.6303666725768\\
7.47	57.6318094028865\\
7.48	57.6332567340224\\
7.49	57.6347086656756\\
7.5	57.6361651974909\\
7.51	57.6376263289946\\
7.52	57.6390920598225\\
7.53	57.6405623895887\\
7.54	57.6420373178028\\
7.55	57.6435168440691\\
7.56	57.6450009679978\\
7.57	57.6464896887608\\
7.58	57.6479830056016\\
7.59	57.6494809182112\\
7.6	57.6509834263318\\
7.61	57.6524905298357\\
7.62	57.6540022275316\\
7.63	57.6555185215338\\
7.64	57.6570394514612\\
7.65	57.6585649732025\\
7.66	57.6600950863029\\
7.67	57.6616297903321\\
7.68	57.6631690849607\\
7.69	57.6647129699251\\
7.7	57.6662614450545\\
7.71	57.6678145104748\\
7.72	57.669372166517\\
7.73	57.6709344134062\\
7.74	57.6725012511343\\
7.75	57.6740726800573\\
7.76	57.6756487009157\\
7.77	57.6772293149417\\
7.78	57.6788145236724\\
7.79	57.6804043287261\\
7.8	57.6819987320222\\
7.81	57.6835977355288\\
7.82	57.6852013410826\\
7.83	57.6868095507347\\
7.84	57.6884223665289\\
7.85	57.6900398283851\\
7.86	57.6916619126338\\
7.87	57.6932886086386\\
7.88	57.6949199183612\\
7.89	57.6965558451791\\
7.9	57.6981963934197\\
7.91	57.6998415655383\\
7.92	57.7014913616456\\
7.93	57.703145780435\\
7.94	57.7048048195889\\
7.95	57.7064684766282\\
7.96	57.7081367490377\\
7.97	57.7098096341374\\
7.98	57.7114871293959\\
7.99	57.7131692310397\\
8	57.7148559353231\\
8.01	57.7165472169545\\
8.02	57.7182430538908\\
8.03	57.7199434343214\\
8.04	57.7216483473061\\
8.05	57.723357783557\\
8.06	57.7250717601008\\
8.07	57.7267902798197\\
8.08	57.7285133098775\\
8.09	57.7302408511369\\
8.1	57.7319729063373\\
8.11	57.7337094806889\\
8.12	57.7354505793175\\
8.13	57.7371962094753\\
8.14	57.7389463829756\\
8.15	57.7407011137674\\
8.16	57.7424604165763\\
8.17	57.7442243072329\\
8.18	57.7459928005641\\
8.19	57.7477659094562\\
8.2	57.7495436459371\\
8.21	57.7513260199722\\
8.22	57.7531130390938\\
8.23	57.7549047098087\\
8.24	57.7567010365434\\
8.25	57.7585020213569\\
8.26	57.7603076652826\\
8.27	57.7621179709547\\
8.28	57.7639329759571\\
8.29	57.7657526206795\\
8.3	57.7675768952355\\
8.31	57.7694057875075\\
8.32	57.7712392852883\\
8.33	57.7730773776475\\
8.34	57.7749200559083\\
8.35	57.7767673129951\\
8.36	57.7786191432251\\
8.37	57.7804755423555\\
8.38	57.7823365067706\\
8.39	57.784202033438\\
8.4	57.7860721202504\\
8.41	57.7879467655675\\
8.42	57.7898259682163\\
8.43	57.7917097277156\\
8.44	57.7935980438639\\
8.45	57.7954909166962\\
8.46	57.7973883466052\\
8.47	57.799290334074\\
8.48	57.80119687975\\
8.49	57.8031080298545\\
8.5	57.8050237564219\\
8.51	57.806944049006\\
8.52	57.8088689127022\\
8.53	57.8107983526753\\
8.54	57.8127323721839\\
8.55	57.814670972281\\
8.56	57.8166141506744\\
8.57	57.8185619025104\\
8.58	57.8205142236137\\
8.59	57.8224711081753\\
8.6	57.824432550932\\
8.61	57.8263985484489\\
8.62	57.8283690996231\\
8.63	57.8303442049508\\
8.64	57.8323238668807\\
8.65	57.8343080896741\\
8.66	57.8362968782076\\
8.67	57.838290237667\\
8.68	57.8402881737199\\
8.69	57.8422906912181\\
8.7	57.8442978182636\\
8.71	57.8463095665005\\
8.72	57.8483258926927\\
8.73	57.8503467923972\\
8.74	57.8523722608717\\
8.75	57.8544022933971\\
8.76	57.8564368858144\\
8.77	57.8584760346404\\
8.78	57.8605197369437\\
8.79	57.8625679906047\\
8.8	57.8646207941736\\
8.81	57.8666781466944\\
8.82	57.8687400479131\\
8.83	57.8708064982098\\
8.84	57.8728774982597\\
8.85	57.8749530491503\\
8.86	57.8770331523808\\
8.87	57.8791178095476\\
8.88	57.8812070225311\\
8.89	57.8833007934994\\
8.9	57.8853991244979\\
8.91	57.8875020198213\\
8.92	57.8896095449954\\
8.93	57.8917216329252\\
8.94	57.8938382849246\\
8.95	57.8959595030443\\
8.96	57.8980852896296\\
8.97	57.9002156473887\\
8.98	57.9023505791263\\
8.99	57.9044900872739\\
9	57.9066341740492\\
9.01	57.9087828414454\\
9.02	57.910936090941\\
9.03	57.913093923713\\
9.04	57.9152563406069\\
9.05	57.917423341839\\
9.06	57.9195949268218\\
9.07	57.9217710924804\\
9.08	57.9239518363579\\
9.09	57.9261371527307\\
9.1	57.9283270346024\\
9.11	57.9305214753042\\
9.12	57.9327204686574\\
9.13	57.9349240344634\\
9.14	57.9371321595745\\
9.15	57.9393448285291\\
9.16	57.9415620387364\\
9.17	57.9437837885193\\
9.18	57.9460100767529\\
9.19	57.9482409028444\\
9.2	57.9504762668608\\
9.21	57.9527161691999\\
9.22	57.9549606105179\\
9.23	57.957209591796\\
9.24	57.9594631140688\\
9.25	57.961721178414\\
9.26	57.9639837860355\\
9.27	57.9662509380718\\
9.28	57.9685226356534\\
9.29	57.9707988799612\\
9.3	57.973079672036\\
9.31	57.9753650128772\\
9.32	57.9776549034786\\
9.33	57.9799493446301\\
9.34	57.9822483370752\\
9.35	57.9845518815512\\
9.36	57.9868599785731\\
9.37	57.9891726286163\\
9.38	57.991489832182\\
9.39	57.9938115895342\\
9.4	57.9961379008963\\
9.41	57.9984687665557\\
9.42	58.0008041864219\\
9.43	58.0031441600062\\
9.44	58.0054886869682\\
9.45	58.0078377670006\\
9.46	58.0101913998371\\
9.47	58.0125495854153\\
9.48	58.014912323524\\
9.49	58.0172796138517\\
9.5	58.019651456178\\
9.51	58.0220278500909\\
9.52	58.0244087950636\\
9.53	58.0267942906934\\
9.54	58.0291843364646\\
9.55	58.0315789317772\\
9.56	58.0339780761567\\
9.57	58.0363817690484\\
9.58	58.038790009811\\
9.59	58.041202797907\\
9.6	58.0436201324102\\
9.61	58.0460420122099\\
9.62	58.0484684367977\\
9.63	58.0508994053851\\
9.64	58.0533349170389\\
9.65	58.055774971206\\
9.66	58.0582195678986\\
9.67	58.0606687062767\\
9.68	58.0631223854402\\
9.69	58.0655806047613\\
9.7	58.0680433637675\\
9.71	58.0705106620211\\
9.72	58.0729824990992\\
9.73	58.0754588745366\\
9.74	58.0779397878509\\
9.75	58.0804252386855\\
9.76	58.0829152267795\\
9.77	58.0854097519883\\
9.78	58.087908814418\\
9.79	58.0904124142878\\
9.8	58.0929205517502\\
9.81	58.0954332266323\\
9.82	58.0979504388472\\
9.83	58.1004721886423\\
9.84	58.1029984766329\\
9.85	58.1055293036665\\
9.86	58.1080646706478\\
9.87	58.1106045786467\\
9.88	58.113149028735\\
9.89	58.1156980218337\\
9.9	58.1182515589144\\
9.91	58.1208096408748\\
9.92	58.123372268529\\
9.93	58.1259394429905\\
9.94	58.1285111653875\\
9.95	58.1310874364196\\
9.96	58.1336682557525\\
9.97	58.1362536226302\\
9.98	58.1388435358737\\
9.99	58.1414379942598\\
10	58.1440369965839\\
10.01	58.1466405415612\\
10.02	58.1492486278377\\
10.03	58.1518612536431\\
10.04	58.1544784171263\\
10.05	58.1571001163891\\
10.06	58.1597263417488\\
10.07	58.1623570858369\\
10.08	58.164992344025\\
10.09	58.1676321116596\\
10.1	58.1702763847749\\
10.11	58.1729251601611\\
10.12	58.1755784356378\\
10.13	58.178236210208\\
10.14	58.1808984835535\\
10.15	58.1835652562734\\
10.16	58.1862365289977\\
10.17	58.188912303331\\
10.18	58.1915925832458\\
10.19	58.1942773743443\\
10.2	58.1969666828649\\
10.21	58.1996605158318\\
10.22	58.2023588803358\\
10.23	58.2050617825929\\
10.24	58.2077692284651\\
10.25	58.2104812232443\\
10.26	58.2131977710212\\
10.27	58.2159188753674\\
10.28	58.2186445392388\\
10.29	58.2213747643357\\
10.3	58.2241095517699\\
10.31	58.2268489020757\\
10.32	58.2295928144982\\
10.33	58.2323412854074\\
10.34	58.2350943104636\\
10.35	58.2378518860285\\
10.36	58.2406140064359\\
10.37	58.2433806654097\\
10.38	58.2461518573405\\
10.39	58.2489275774629\\
10.4	58.2517078218257\\
10.41	58.2544925874957\\
10.42	58.2572818722223\\
10.43	58.2600756740877\\
10.44	58.2628739916834\\
10.45	58.2656768239931\\
10.46	58.2684841702077\\
10.47	58.271296029916\\
10.48	58.2741124030168\\
10.49	58.2769332895454\\
10.5	58.2797586897608\\
10.51	58.2825886040514\\
10.52	58.285423032826\\
10.53	58.2882619765992\\
10.54	58.2911054349952\\
10.55	58.2939534102279\\
10.56	58.2968059052586\\
10.57	58.2996629229798\\
10.58	58.3025244657497\\
10.59	58.3053905348482\\
10.6	58.3082611285718\\
10.61	58.3111362437961\\
10.62	58.3140158766146\\
10.63	58.3169000223405\\
10.64	58.3197886768475\\
10.65	58.3226818375182\\
10.66	58.3255795033692\\
10.67	58.3284816751711\\
10.68	58.3313883548467\\
10.69	58.3342995451463\\
10.7	58.3372152496547\\
10.71	58.3401354723642\\
10.72	58.3430602173222\\
10.73	58.3459894890675\\
10.74	58.3489232896065\\
10.75	58.3518616178175\\
10.76	58.3548044712059\\
10.77	58.3577518464009\\
10.78	58.36070373991\\
10.79	58.3636601483758\\
10.8	58.3666210688514\\
10.81	58.3695864990677\\
10.82	58.3725564373074\\
10.83	58.3755308822953\\
10.84	58.3785098333226\\
10.85	58.381493290125\\
10.86	58.384481252762\\
10.87	58.3874737217357\\
10.88	58.3904706978295\\
10.89	58.3934721819063\\
10.9	58.3964781750191\\
10.91	58.3994886782935\\
10.92	58.4025036927652\\
10.93	58.4055232195377\\
10.94	58.4085472596551\\
10.95	58.4115758132953\\
10.96	58.4146088802917\\
10.97	58.417646461073\\
10.98	58.420688556339\\
10.99	58.4237351671275\\
11	58.4267862947229\\
};
\addlegendentry{Safe Imitation Learning}

\addplot [color=red, dashed, line width=1.0pt]
  table[row sep=crcr]{%
0	60.75\\
0.260385036468506	60.4375\\
0.514420747756958	60.3125\\
0.77122974395752	60.125\\
1.01579904556274	59.9375\\
1.25869584083557	59.875\\
1.51554679870605	59.75\\
1.76970887184143	59.75\\
2.01346015930176	59.8125\\
2.26456522941589	59.75\\
2.51743507385254	59.5\\
2.76477599143982	59.3125\\
3.01760363578796	59.1875\\
3.26084804534912	58.9375\\
3.51035284996033	58.6875\\
3.76422071456909	58.4375\\
3.97685432434082	58.25\\
4.25904893875122	58.1875\\
4.51348280906677	58.375\\
4.72662687301636	57.9375\\
5.01741242408752	57.6875\\
5.26057887077332	57.5625\\
5.49404287338257	57.625\\
5.76330590248108	57.4375\\
6.01699709892273	57.4375\\
6.24035453796387	57.375\\
6.5173761844635	57.375\\
6.77055478096008	57.375\\
6.98401379585266	57.5625\\
7.26666951179504	57.5\\
7.52012586593628	57.375\\
7.72354412078857	57.25\\
8.0179078578949	57.25\\
8.27118420600891	57.4375\\
8.52349710464478	57.5\\
8.76748514175415	57.6875\\
9.02075958251953	57.8125\\
9.27296495437622	58.1875\\
9.52661061286926	58.4375\\
9.77027153968811	58.9375\\
10.0201983451843	59.0625\\
10.2638461589813	59.125\\
10.5174899101257	59.25\\
10.7711069583893	59.3125\\
11.0141401290894	59.6875\\
};
\addlegendentry{Human Driving}

\end{axis}

\end{tikzpicture}%

%% file: images/safe.tex
%
%
\definecolor{mycolor1}{rgb}{0.46667,0.67451,0.18824}%
\definecolor{mycolor2}{rgb}{0.85098,0.32549,0.09804}%
\begin{tikzpicture}

\begin{axis}[%
width=0.95\figurewidth,
height=1.5\figureheight,
legend cell align={center},
legend style={
	fill opacity=0.8,
	draw opacity=1,
	text opacity=1,
	at={(0.1,-0.55)},
	anchor=south west,
	draw=white!80!black
},
scale only axis,
bar shift auto,
xmin=-0.2,
xmax=11.2,
xtick={ 1,  2,  3,  4,  5,  6,  7,  8,  9, 10},
xlabel style={font=\color{white!15!black}},
xlabel={Test Scenario \#},
ymin=0,
ymax=1,
title={Safety Performance},
ylabel style={font=\color{white!15!black}},
ylabel={Path Completed [\%]},
axis background/.style={fill=white},
]
\addplot[ybar, bar width=5, fill=mycolor1, draw=black, area legend] table[row sep=crcr] {%
1	1\\
2	1\\
3	1\\
4	0.243652537072041\\
5	1\\
6	0.251206141365427\\
7	0.286587934750144\\
8	1\\
9	1\\
10	1\\
};
\addlegendentry{Safe Imitation Learning}

\addplot[ybar, bar width=5, fill=mycolor2, draw=black, area legend] table[row sep=crcr] {%
	1	0.305120983855612\\
	2	0.085567405490334\\
	3	0.24590380413413\\
	4	0.188162182557627\\
	5	0.386465200991958\\
	6	0.374438391522503\\
	7	0.121543460031289\\
	8	0.512563164392433\\
	9	0.529482717115704\\
	10	0.442272382240267\\
};
\addlegendentry{Behavioral Cloning}

\addplot[forget plot, color=white!15!black] table[row sep=crcr] {%
-0.2	0\\
11.2	0\\
};

\end{axis}
\end{tikzpicture}%